\documentclass[prd,twocolumn,eqsecnum,superscriptaddress,showpacs,letterpaper,altaffilletter]{revtex4}
\usepackage{color}
\usepackage{times}
\usepackage{graphicx}
\usepackage{fancyhdr}
\usepackage{latexsym}
\usepackage{float}

\def\thercsid{\relax}
\def\rcsid#1{\def\next##1#1{\def\thercsid{##1}}\next}
\rcsid$Revision: 1.23 $

\renewcommand{\today}{\number\day\space\ifcase\month\or
  January\or February\or March\or April\or May\or June\or
  July\or August\or September\or October\or November\or December\fi
  \space\number\year}

\def\etc{{\it etc.}}
\def\eg{{\it e.g.}}

\def\ie{{\it i.e.}}

\def\be{\begin{equation}}
\def\ee{\end{equation}}
\def\bi{\begin{itemize}} 
\def\ei{\end{itemize}}
\def\ben{\begin{enumerate}}
\def\een{\end{enumerate}}
\def\hrssu{Hz${}^{-1/2}$}
\def\hrss{h_\mathrm{rss}}
\newcommand\ligodoc{P040040-07-R}

\begin{document}

\title{
Upper limits on gravitational wave bursts in LIGO's second science run
\\
{\color{red} {\large LIGO-\ligodoc} \\ 
to be submitted for publication to Phys Rev D} }

\begin{abstract}  
\vspace*{0.2in}

We perform a search for gravitational wave bursts using data from
the second science run of the LIGO detectors,
using a method based on
a wavelet time-frequency decomposition.
This search is sensitive to bursts of duration much less than a second
and with frequency
content in the 100--1100~Hz range.
It features significant improvements in the instrument sensitivity
and in the analysis pipeline
with respect to the burst search previously reported by LIGO.
Improvements in the search method allow exploring weaker signals,
relative to the detector noise floor,
while maintaining
a low false alarm rate, $O(0.1)$~$\mu$Hz.
The sensitivity in terms of the {\it root-sum-square} (rss) strain amplitude
lies in the range of
$\hrss \sim 10^{-20} - 10^{-19}$~\hrssu.
No gravitational wave signals were detected in 9.98 days of analyzed data.
We interpret the search result in terms of a frequentist upper limit
on the rate of detectable gravitational wave bursts at the level of 0.26
events per day at 90\% confidence level.
We combine this limit with measurements of the detection efficiency
for given waveform morphologies in order to yield rate versus strength
exclusion curves as well as to establish order-of-magnitude
distance sensitivity to certain modeled astrophysical sources.
Both the rate upper limit and its applicability to signal strengths
improve our previously reported limits and reflect
the most sensitive broad-band search
for untriggered and unmodeled gravitational wave bursts to date.
\end{abstract}


\date[\relax]{ RCS \thercsid; compiled \today }

\newcommand*{\AG}{Albert-Einstein-Institut, Max-Planck-Institut f\"ur Gravitationsphysik, D-14476 Golm, Germany}

\affiliation{\AG}

\newcommand*{\AH}{Albert-Einstein-Institut, Max-Planck-Institut f\"ur Gravitationsphysik, D-30167 Hannover, Germany}

\affiliation{\AH}

\newcommand*{\AN}{Australian National University, Canberra, 0200, Australia}

\affiliation{\AN}

\newcommand*{\CH}{California Institute of Technology, Pasadena, CA  91125, USA}

\affiliation{\CH}

\newcommand*{\DO}{California State University Dominguez Hills, Carson, CA  90747, USA}

\affiliation{\DO}

\newcommand*{\CA}{Caltech-CaRT, Pasadena, CA  91125, USA}

\affiliation{\CA}

\newcommand*{\CU}{Cardiff University, Cardiff, CF2 3YB, United Kingdom}

\affiliation{\CU}

\newcommand*{\CL}{Carleton College, Northfield, MN  55057, USA}

\affiliation{\CL}

\newcommand*{\FN}{Fermi National Accelerator Laboratory, Batavia, IL  60510, USA}

\affiliation{\FN}

\newcommand*{\HC}{Hobart and William Smith Colleges, Geneva, NY  14456, USA}

\affiliation{\HC}

\newcommand*{\IU}{Inter-University Centre for Astronomy  and Astrophysics, Pune - 411007, India}

\affiliation{\IU}

\newcommand*{\CT}{LIGO - California Institute of Technology, Pasadena, CA  91125, USA}

\affiliation{\CT}

\newcommand*{\LM}{LIGO - Massachusetts Institute of Technology, Cambridge, MA 02139, USA}

\affiliation{\LM}

\newcommand*{\LO}{LIGO Hanford Observatory, Richland, WA  99352, USA}

\affiliation{\LO}

\newcommand*{\LV}{LIGO Livingston Observatory, Livingston, LA  70754, USA}

\affiliation{\LV}

\newcommand*{\LU}{Louisiana State University, Baton Rouge, LA  70803, USA}

\affiliation{\LU}

\newcommand*{\LE}{Louisiana Tech University, Ruston, LA  71272, USA}

\affiliation{\LE}

\newcommand*{\LL}{Loyola University, New Orleans, LA 70118, USA}

\affiliation{\LL}

\newcommand*{\MP}{Max Planck Institut f\"ur Quantenoptik, D-85748, Garching, Germany}

\affiliation{\MP}

\newcommand*{\MS}{Moscow State University, Moscow, 119992, Russia}

\affiliation{\MS}

\newcommand*{\ND}{NASA/Goddard Space Flight Center, Greenbelt, MD  20771, USA}

\affiliation{\ND}

\newcommand*{\NA}{National Astronomical Observatory of Japan, Tokyo  181-8588, Japan}

\affiliation{\NA}

\newcommand*{\NO}{Northwestern University, Evanston, IL  60208, USA}

\affiliation{\NO}

\newcommand*{\SC}{Salish Kootenai College, Pablo, MT  59855, USA}

\affiliation{\SC}

\newcommand*{\SE}{Southeastern Louisiana University, Hammond, LA  70402, USA}

\affiliation{\SE}

\newcommand*{\SA}{Stanford University, Stanford, CA  94305, USA}

\affiliation{\SA}

\newcommand*{\SR}{Syracuse University, Syracuse, NY  13244, USA}

\affiliation{\SR}

\newcommand*{\PU}{The Pennsylvania State University, University Park, PA  16802, USA}

\affiliation{\PU}

\newcommand*{\TC}{The University of Texas at Brownsville and Texas Southmost College, Brownsville, TX  78520, USA}

\affiliation{\TC}

\newcommand*{\TR}{Trinity University, San Antonio, TX  78212, USA}

\affiliation{\TR}

\newcommand*{\HU}{Universit{\"a}t Hannover, D-30167 Hannover, Germany}

\affiliation{\HU}

\newcommand*{\BB}{Universitat de les Illes Balears, E-07122 Palma de Mallorca, Spain}

\affiliation{\BB}

\newcommand*{\BR}{University of Birmingham, Birmingham, B15 2TT, United Kingdom}

\affiliation{\BR}

\newcommand*{\FA}{University of Florida, Gainesville, FL  32611, USA}

\affiliation{\FA}

\newcommand*{\GU}{University of Glasgow, Glasgow, G12 8QQ, United Kingdom}

\affiliation{\GU}

\newcommand*{\MU}{University of Michigan, Ann Arbor, MI  48109, USA}

\affiliation{\MU}

\newcommand*{\OU}{University of Oregon, Eugene, OR  97403, USA}

\affiliation{\OU}

\newcommand*{\RO}{University of Rochester, Rochester, NY  14627, USA}

\affiliation{\RO}

\newcommand*{\UW}{University of Wisconsin-Milwaukee, Milwaukee, WI  53201, USA}

\affiliation{\UW}

\newcommand*{\WU}{Washington State University, Pullman, WA 99164, USA}

\affiliation{\WU}

\author{B.~Abbott}    \affiliation{\CT}

\author{R.~Abbott}    \affiliation{\LV}

\author{R.~Adhikari}    \affiliation{\LM}

\author{A.~Ageev}    \affiliation{\MS}  \affiliation{\SR}

\author{B.~Allen}    \affiliation{\UW}

\author{R.~Amin}    \affiliation{\FA}

\author{S.~B.~Anderson}    \affiliation{\CT}

\author{W.~G.~Anderson}    \affiliation{\TC}

\author{M.~Araya}    \affiliation{\CT}

\author{H.~Armandula}    \affiliation{\CT}

% ...

 \collaboration{The LIGO Scientific Collaboration, http://www.ligo.org}

 \noaffiliation

\maketitle

\section{Introduction}\label{sec:introduction}

The Laser Interferometer Gravitational wave Observatory (LIGO)
is a network of
interferometric detectors
aiming to make direct observations of gravitational
waves.
Construction of the LIGO detectors is essentially complete, and
much progress has been made in commissioning them to
(a) bring the three interferometers to their
final optical configuration, (b) reduce the interferometers' noise
floors and improve the stationarity of the noise, and
(c) pave the way toward long-term science observations.
Interleaved with commissioning, four ``science runs'' have been
carried out to collect data under stable operating conditions for
astrophysical gravitational wave searches, albeit at reduced
sensitivity and observation time relative to the LIGO design goals.
The first science run, called S1, took place in the summer of 2002
over a period of 17 days.
S1 represented a major milestone as the longest and most sensitive operation
of broad-band interferometers {\em in coincidence} up to that time.
Using the S1 data from the LIGO and GEO600
interferometers~\cite{ref:LIGOS1instpaper},
astrophysical searches for four general categories
of gravitational wave source types---binary inspiral~\cite{ref:IUL},
burst-like~\cite{ref:BUL}, stochastic~\cite{ref:SUL} and
continuous wave~\cite{ref:PUL}---were pursued by the LIGO Scientific
Collaboration (LSC).
These searches established general methodologies to be followed and
improved upon for the analysis of data from future runs.
In 2003 two additional science runs of the LIGO instruments
collected data of improved sensitivity with respect to S1,
but still less sensitive than the instruments' design goal.
The second science run (S2) collected data in early 2003 and
the third science run (S3) at the end of the same year.
Several searches have been completed or are underway using data from
the S2 and S3 runs~\cite{ref:S2knownpulsar,ref:GRB030329,ref:IUL2,ref:S2Macho,ref:tamaligos2,ref:S2hough,ref:S2BBH,ref:S3stoch}.
A fourth science run, S4, took place at the beginning of 2005.

In this paper we report the results of a search for
gravitational wave bursts using the LIGO S2 data.
The astrophysical motivation for burst events is strong;
it embraces catastrophic phenomena in the universe 
with or without clear signatures in the electromagnetic spectrum like
supernova explosions~\cite{ref:ZM,ref:DFM,ref:OBLW},
the merging of compact binary stars
as they form a single black hole~\cite{ref:Laz,lazarus,ref:FlanHughes}
and the astrophysical engines that power the
gamma ray bursts~\cite{ref:grb}.
Perturbed or accreting black holes, neutron star oscillation modes and
instabilities as well as cosmic string cusps and
kinks~\cite{ref:strings04}
are also potential burst sources.
The expected rate, strength and waveform morphology
for such events is not generally known.
For this reason, our assumptions for the expected signals
are minimal.
The experimental signatures on which this search focused 
can be described as burst signals of short duration ($\ll$ 1~second)
and  with enough
signal strength in the LIGO sensitive band (100--1100~Hz)
to be detected
in coincidence in all three LIGO instruments.
The triple coincidence requirement is used to reduce the false alarm
rate (background) to much less than one event over the course of the
run, so that even a single event candidate would have high statistical
significance.

The general methodology in pursuing this search follows the
one we presented in the analysis of the S1 data~\cite{ref:BUL}
with some significant improvements.
In the S1 analysis the ringing of the pre-filters limited
our ability to perform tight time-coincidence between the
triggers coming from the three LIGO instruments.
This is addressed
by the use of a new search method that does not require strong
pre-filtering.
This new method also provides an improved event parameter estimation,
including timing resolution.
Finally, a waveform consistency test is introduced for events that
pass the time and frequency coincidence requirements in the
three LIGO detectors.

This search examines 9.98 days of live time 
and yields one candidate event in coincidence
among the three LIGO detectors during S2.
Subsequent examination of this event reveals an acoustic origin for the
signal in the two Hanford detectors,
easily eliminated using a ``veto'' based on acoustic power in a microphone. 
Taking this into account, we set an upper limit on the
rate of burst events detectable by our detectors at the level of 0.26 per day
at an estimated 90\% confidence level.
We have used {\it{ad hoc}} waveforms (sine-Gaussians and Gaussians)
to establish the sensitivity
of the S2 search pipeline and to interpret our upper limit as an
excluded region in the space of signal rate versus strength.
The burst search sensitivity in terms of the
{\it root-sum-square} (rss) strain amplitude incident on Earth
lies in the range $\hrss \sim 10^{-20} - 10^{-19}$~\hrssu.
Both the upper limit (rate) and its applicability to signal strengths
(sensitivity)
reflect significant improvements with respect to our S1 result~\cite{ref:BUL}.
In addition, we evaluate the sensitivity of the search to
astrophysically motivated waveforms derived from models of stellar
core collapse~\cite{ref:ZM,ref:DFM,ref:OBLW} and from the merger of binary
black holes~\cite{ref:Laz,lazarus}. 

In the following sections we describe the LIGO instruments and the
S2 run in more detail (section~\ref{sec:ligos2run}) as well as
an overview of the search pipeline (section~\ref{sec:pipeline}).
The procedure for selecting the data that we analyze is described
in section~\ref{sec:dataselect}.
We then present the search algorithm and the waveform consistency test
used in the event selection (section~\ref{sec:etg_methods})
and discuss the role of vetoes in this search (section~\ref{sec:vetoes}).
Section~\ref{sec:evana} describes the final event analysis and the
assignment of an upper limit on the rate of detectable bursts.
The efficiency of the search for various target waveforms is
presented in section~\ref{sec:efficiency}.
Our final results and discussion are presented in sections~\ref{sec:results}
and~\ref{sec:summaries}.

\section{The Second LIGO Science Run}\label{sec:ligos2run}

LIGO comprises three interferometers at two sites: an interferometer
with 4~km long arms at the LIGO Livingston Observatory in Louisiana
(denoted L1) and interferometers with 4~km and 2~km long arms in a
common vacuum system at the LIGO Hanford Observatory in Washington
(denoted H1 and H2).
All are Michelson interferometers with power
recycling and resonant cavities in the two arms to increase the
storage time (and consequently the phase shift) for the light
returning to the beam splitter due to motions of the end
mirrors~\cite{ref:Saulson:1994}.
The mirrors are suspended as pendulums from vibration-isolated
platforms to protect them from external noise sources.  A detailed
description of the LIGO detectors as they were configured for the
S1 run may be found in ref.~\cite{ref:LIGOS1instpaper}.

\subsection{Improvements to the LIGO detectors for S2}

The LIGO interferometers~\cite{ref:LIGOS1instpaper,siggCQG}
are still undergoing commissioning and
have not yet reached their final operating configuration and
sensitivity.  Between S1 and S2 a number of changes were made
which resulted in improved sensitivity as well as overall
instrument stability and stationarity.
The most important of these are summarized below.

The mirrors' analog suspension controller electronics on the H2 and L1
interferometers were replaced with digital controllers
of the type installed on H1 before the S1 run. The addition of a
separate DC bias supply for alignment relieved the range
requirement of the suspensions' coil drivers.  This, combined with flexibility
of a digital system capable of coordinated switching of analog and
digital filters, enabled the new coil drivers to operate with much
lower electronics output noise.  In particular, the system had two
separate modes of operation: acquisition mode with larger range
and noise, and run mode with reduced range and noise.
A matched pair of filters was used to minimize noise in the coil
current due to the discrete steps in the digital to analog converter
(DAC) at the output of the digital suspension controller: a digital
filter before the DAC boosted the high-frequency component relative
to the low-frequency component, and an analog filter after the DAC
restored their relative amplitudes.
Better filtering, better diagonalization of the drive to the coils to
eliminate length-to-angle couplings and more flexible
control/sequencing features also contributed to an overall
performance improvement.

The noise from the optical lever servos that damp the angular
excitations of the interferometer optics was reduced.  The
mechanical support elements for the optical transmitter and
receiver were stiffened to reduce low frequency
vibrational excitations. 
Input noise to
the servo due to the discrete steps in the analog to digital
converter (ADC) was reduced by a filter pair surrounding the
ADC: an analog filter to whiten the data going into the ADC and a
digital filter to restore it to its full dynamic range.

Further progress was made on commissioning the wavefront sensing
(WFS) system for alignment control of the H1 interferometer. This
system uses the main laser beam to sense the proper alignment for
the suspended optics. During S1, all interferometers had two
degrees of freedom for the main interferometer (plus four degrees
of freedom for the mode cleaner) controlled by their WFS.  For S2,
the H1 interferometer had 8 out of 16 alignment degrees of freedom for
the main interferometer under WFS control.  As a result, it
maintained a much more uniform operating point over the run than
the other two interferometers, which continued to have only two
degrees of freedom under WFS control.

The high frequency sensitivity was increased by operating the
interferometers with higher effective power.  Two main factors
enabled this power increase.  Improved alignment techniques and
better alignment stability (due to the optical lever and
wavefront sensor improvements described above) reduced the amount of spurious
light at the anti-symmetric port, which would have saturated the
photodiode if the laser power had been increased in S1.  Also,
a new servo system to cancel the out-of-phase (non-signal) photocurrent
in the anti-symmetric photodiode was added.  This amplitude of the
out-of-phase photocurrent is nominally zero for a perfectly aligned
and matched interferometer, but various imperfections in the
interferometer can lead to large low frequency signals.  The new
servo prevents these signals from causing saturations in the
photodiode and its RF preamplifier.  During S2, the
interferometers operated with about $1.5$~W incident on the
mode cleaner and about $40$~W incident on the beam splitter.

\begin{figure}[!thb]
\includegraphics[width=0.99\linewidth]{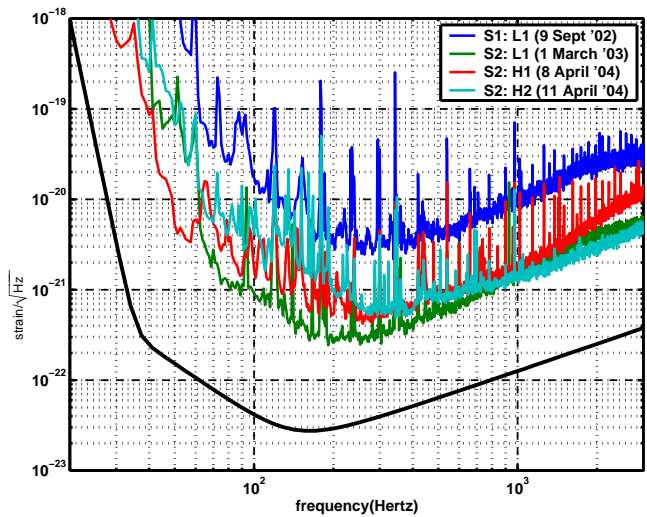}
\caption{Typical LIGO strain sensitivities in units of~\hrssu\
during the second science run (S2), compared to the most sensitive
detector (L1) during the S1 science run. The solid line denotes the design
goal for the 4~km instruments.
}
\label{fig:s2sens}
\end{figure}

These changes led to a significant improvement in detector
sensitivity.  Figure~\ref{fig:s2sens} shows typical spectra
achieved by the LIGO
interferometers during the S2 run compared with LIGO's S1 and design
sensitivity.
The differences among the three LIGO S2
spectra reflect differences in the operating parameters and
hardware implementations of the three instruments, which were in
various stages of reaching the final design configuration.

\subsection{Data from the S2 run}

The data analyzed in this paper were taken during LIGO's second
science run (S2), which spanned 59 days from
February 14 to April 14, 2003. During this time, operators and
scientific monitors worked to maintain continuous low noise
operation of the LIGO instruments.
The duty cycles for the individual interferometers, defined as the
fraction of the total run time when the interferometer was locked
(\ie, all interferometer control servos
operating in their linear regime)
and in its low noise configuration, 
were approximately
74\% for H1, 58\% for H2 and 37\% for L1;
the triple coincidence duty cycle (\ie, the time during which
all three interferometers were simultaneously
in lock and in low-noise configuration) was 22\%.
The longest continuous locked
stretch for any interferometer during S2 was 66.2 hours for H1.
The main sources of lost time were high microseismic motion at both
sites due to storms, and anthropogenic noise in the vicinity of the Livingston
Observatory.

Improved monitoring and automated alarms instituted after S1 gave
the operators and scientific monitors better warnings of
out-of-nominal operating conditions for the interferometers.  As a
result, the fraction of time lost to high noise or to missing
calibration lines (both major sources of unanalyzable data during
the S1 run) was greatly reduced.  Thus, even though the S2
run was less than a factor of four longer than the S1 run and the
duty cycle for triple interferometer coincidence was in fact
marginally lower (23.4\% for S1 vs. 22.0\% for S2), the total
amount of {\em analyzable} triple coincidence data was 305 hours
compared to 34 hours for S1.

The signature of a gravitational wave is a differential change in the
lengths of the two interferometer arms relative to the nominal
lengths established by the control system, $s(t) = [\Delta L_x(t) -
\Delta L_y(t)]/L$ , where $L$ is the average length of the $x$ and $y$ arms.
As in S1, this time series was derived from the error signal of the
feedback loop used to differentially control the lengths of the
interferometer arms in order to keep the optical cavities on resonance.
To calibrate the error signal, the effect of
the feedback loop gain was measured and divided out. Although more
stable than during S1, the response functions varied over the
course of the S2 run due to drifts in the alignment of the optical
elements. These were tracked by injecting fixed-amplitude sinusoidal
signals (calibration lines) into the differential arm control
loop, and monitoring the amplitudes of these signals at the
measurement (error) point~\cite{ref:S2calibration}.

The S2 run also involved coincident running with the TAMA
interferometer~\cite{ref:TAMAref}. TAMA achieved a duty cycle of 81\%
and had a sensitivity comparable to LIGO's above
$\sim 1$~kHz,
but had poorer sensitivity at lower frequencies where the LIGO
detectors had their best sensitivity.  In addition, the location
and orientation of the TAMA detector differs substantially from
the LIGO detectors, which further reduced the chance of a
coincident detection at low frequencies.  For these reasons, the
joint analysis of LIGO and TAMA data focused on gravitational wave
frequencies from 700--2000~Hz
and will be described in a separate paper~\cite{ref:tamaligos2}.
In this paper, we report the result of a LIGO-only search for signals
in the range 100--1100~Hz.
The overlap between these two searches
(700--1100~Hz) serves to ensure that possible sources with
frequency content spanning the two searches will not be missed.
The GEO600 interferometer~\cite{ref:GEOref}, which collected data
simultaneously with LIGO during the S1 run, was undergoing
commissioning at the time of the S2 run.

\section{Search Pipeline Overview}\label{sec:pipeline}

The overall burst search pipeline used in the S2 analysis follows
the one we introduced in our S1 search~\cite{ref:BUL}.
First, data selection criteria are applied
in order to define periods when
the instruments are well behaved and the recorded data can be used
for science searches (section~\ref{sec:dataselect}).

A wavelet-based algorithm called WaveBurst~\cite{ref:wbmethod2,ref:wbmethod} 
(section~\ref{sec:etg_methods}) is then used to identify
candidate burst events.
Rather than operating on the data from a single interferometer,
WaveBurst analyzes simultaneously the time series coming from a pair
of interferometers and incorporates strength thresholding as well as
time and frequency coincidence to identify transients with consistent
features in the two data streams.
To reduce the false alarm rate, we further require that
candidate gravitational wave events occur effectively simultaneously in
all three LIGO detectors (section~\ref{ss:triplecoinc}).
Besides requiring compatible WaveBurst event parameters,
this involves a waveform consistency
test, the $r$-statistic~\cite{rstatCQG} (section~\ref{sec:rstat}), which
is based on forming the normalized linear correlation of the
raw time series coming from the LIGO instruments.
This test takes advantage of the fact that the
arms of the interferometers at the
two LIGO sites are nearly co-aligned, and therefore a gravitational
wave generally will produce correlated time series.
The use of WaveBurst and the $r$-statistic are the major changes
in the S2 pipeline with respect to the pipeline used for S1~\cite{ref:BUL}.

When candidate burst events are identified,
they can be checked against veto conditions based on the many
auxiliary read-back channels of the servo control systems and
physical environment monitoring channels that are recorded in the LIGO
data stream (section~\ref{sec:vetoes}).

The background in this search is measured by
artificially shifting in time the raw time series of
one of the LIGO instruments, L1, and repeating the analysis as for
the un-shifted data.
The time-shifted case will often be referred to as ``time-lag'' data 
and the unshifted case as ``zero-lag'' data.
We will describe the background estimation in more detail in
section~\ref{sec:evana}.

We have relied on hardware and software signal ``injections'' in order to establish
the efficiency of the pipeline.
Simulated signals with various morphologies~\cite{ref:S2MDC} were
added to the digitized raw data time series at the beginning of our
analysis pipeline and were
used to establish the fraction of detected events as
a function of their strength (section~\ref{sec:efficiency}).
The same analysis pipeline was used to analyze raw (zero-lag), time-lag,
and injection data samples.

We maintain a detailed list with a number of checks to perform for any
zero-lag event(s) surviving the analysis pipeline to evaluate
whether they could plausibly be gravitational wave bursts.
This ``detection check-list'' is updated as
we learn more about the instruments and refine our
methodology.
A major aspect is the
examination of environmental and auxiliary interferometric
channels in order to identify terrestrial disturbances
that might produce a candidate event through some coupling mechanism.
Any remaining events are compared with the background and the
experiment's live time in order to establish a detection or
an upper limit on the
rate of burst events.

\section{Data Selection}\label{sec:dataselect}

The selection of data to be analyzed was a key first step in this search.
We expect a gravitational wave to appear in all three LIGO instruments,
although in some cases it may be at or below the level of the noise.
For this search, we {\em require} a signal above the noise baseline
in all three instruments in order to suppress the rate of noise
fluctuations that may fake astrophysical burst events.
In the case of a genuine astrophysical event this requirement
will not only increase our detection confidence but it will also
allow us to extract in the best possible way the signal and source parameters.
Therefore, for this search we have confined ourselves to periods of
time when all three LIGO interferometers were simultaneously locked in
low noise mode with nominal operating parameters (nominal servo loop gains,
nominal filter settings, etc.), marked by a manually set bit
(``science mode'') in the data stream.
This produced a total of
318 hours of potential data for analysis.  This total was reduced
by the following data selection cuts:

\begin{itemize}

\item A minimum duration of 300 seconds was required for a
triple coincidence segment to be analyzed for this search. This cut eliminated
$0.9$\% of the initial data set.

\item Post-run re-examination of the interferometer configuration
and status channels included in the data stream identified a small
amount of time when the interferometer configuration deviated from
nominal.
In addition we identified short periods of time when the timing system
for the data acquisition had lost synchronization.  
These cuts reduced the data set by $0.2$\%.

\item It was discovered that large low frequency excitations of
the interferometer could cause the photodiode at the
anti-symmetric port to saturate.  This caused bursts of excess
noise due to nonlinear up-conversion.  These periods of time were
identified and eliminated, reducing the data set by 0.3\%.

\item There were occasional periods of time when the calibration
lines
either were absent or were significantly weaker than normal. 
Eliminating these periods reduced the data set by approximately 2\%.

\item The H1 interferometer had a known problem with a marginally
stable servo loop, which occasionally led to higher than normal noise in the
error signal for the differential arm length (the channel used in this
search for gravitational waves).
A data cut was imposed to eliminate periods of time when 
the RMS noise in the 200--400~Hz band of
this channel exceeded a threshold value for 5 consecutive minutes.
The requirement for 5 consecutive minutes was imposed to prevent a
short burst of gravitational waves (the object of this search)
from triggering this cut.
This cut reduced the data set by 0.4\%.

\end{itemize}

These data quality cuts eliminated a total of 13 hours from the
original 318 hours of triple coincidence data, leaving a ``live-time''
of 305 hours.
The fraction of
data surviving these quality cuts (96\%) is a significant
improvement over the experience in S1 when only 37\% of the data
passed all the quality cuts.

The trigger generation software used in this search (to be described
in the next section) processed data in fixed 2-minute time intervals,
requiring good data quality for the entire interval.  This constraint,
along with other constraints imposed by other trigger generation
methods which were initially used to define a common data set, led to
a net loss of 41 hours, leaving 264 hours of triple coincidence data
actually searched.

The search for bursts in the LIGO S2 data used roughly 10\% of the
triple coincidence data set in order to tune the
pipeline (as described below) and establish event selection criteria.
This data set was chosen uniformly across the acquisition time and
constituted the so-called ``playground'' for the search.
The rate bound calculated in Sec.~\ref{sec:evana} reflects only the
remaining $\sim$90\% of the data, in order to avoid bias from the tuning
procedures.

\section{Methods for Event Trigger Selection}\label{sec:etg_methods}

An accurate knowledge of gravitational wave burst waveforms would
allow the use of {\it{matched filtering}}~\cite{WainsteinZubakov}
along the lines of the
search for binary inspirals~\cite{ref:IUL,ref:IUL2}.  However, many
different astrophysical systems may give rise to gravitational wave
bursts, and the physics of these systems is often very complicated.
Even when numerical relativistic calculations have been carried out,
as in the case of core collapse supernovae, they generally yield
roughly representative waveforms rather than exact predictions.
Therefore, our present searches for gravitational wave bursts use general
algorithms which are sensitive to a wide range of potential signals.

The first LIGO burst search~\cite{ref:BUL}
used two Event Trigger Generator (ETG)
algorithms: a time-domain method designed to detect a large ``slope''
(time derivative) in the data stream after
suitable filtering~\cite{arnaud99a,pradier01a}, and a
method called {\sc tfclusters}~\cite{sylvestre02b} which is based on
identifying clusters
of excess power in time-frequency spectrograms.
Several other burst-search methods have been developed by members of
the LIGO Scientific Collaboration.  For this paper, we have chosen to
focus on a single ETG called WaveBurst which identifies clusters of
excess power once the signal is decomposed in the wavelet domain, as
described below.  Other methods which were applied to the S2 data
include {\sc tfclusters}; the excess power statistic
of Anderson {\it et al.}~\cite{ref:power};
and the
``Block-Normal'' time-domain algorithm~\cite{BNCQG}.
In preliminary studies using S2 playground data, these other methods
had sensitivities comparable to WaveBurst for
the target waveforms described in section~\ref{sec:efficiency},
but their implementations were less mature at the time of this analysis.

An integral part of our S2 search and the final event trigger selection
is to perform a consistency test among the data streams recorded by
the different interferometers at each trigger time identified by the ETG.
This is done using the $r$-statistic~\cite{rstatCQG},
a time-domain cross-correlation method sensitive to the coherent part
of the candidate signals, described in subsection C below.

\subsection{WaveBurst}\label{sec:etg_methods_WaveBurst}

WaveBurst is an ETG
that searches for gravitational wave bursts in the wavelet
time-frequency domain.
It is described in greater detail in~\cite{ref:wbmethod2,ref:wbmethod}.
The method uses wavelet transformations in order to obtain the
time-frequency representation of the data.
Bursts are identified by searching for regions 
in the wavelet time-frequency domain with an excess of power,
coincident between two or more interferometers,
that is inconsistent with stationary detector noise. 

\begin{figure}[!thb]
\includegraphics[width=0.98\linewidth]{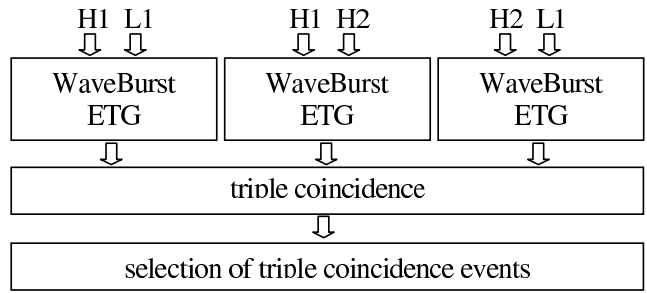}
\caption{Block diagram of the WaveBurst analysis pipeline for the
  three LIGO detectors, H1, H2 and L1 as applied in the S2 data.}
\label{fig:pipeline}
\end{figure}

WaveBurst processes gravitational wave data from two interferometers
at a time. As shown in Fig.~\ref{fig:pipeline} 
the analysis is 
performed over three LIGO detectors resulting in the production of triggers 
for three detector pairs. 
The three sets of triggers are then compared in a ``triple
coincidence'' step which checks for consistent trigger times and
frequency components, as will be described in
Section~\ref{ss:triplecoinc}.

For each detector pair, the WaveBurst ETG performs the following steps:
(a) wavelet transformation applied to the gravitational wave channel
from each detector,
(b) selection of wavelet amplitudes exceeding a threshold,
(c) identification of common wavelet components in the two channels,
(d) clustering of nearby wavelet components, and
(e) selection of burst triggers. 
During steps (a), (b) and (d) the data processing is independent for
each channel. 
During  steps (c) and (e) data from both channels are used.

The input data to the WaveBurst ETG are
time series from the gravitational wave channel
with duration of $120$ seconds and
sampling rate of $16384$~Hz. Before the wavelet transformation is applied
the data are downsampled by a factor of two. Using an orthogonal wavelet   
transformation (based on a symlet
wavelet with filter length of 60)
the time series are converted into wavelet series $W_{ij}$, where
$i$ is the time index and $j$ is the wavelet layer index. Each wavelet
layer can be associated with a certain frequency band of the initial time 
series.
The time-frequency resolution of the WaveBurst scalograms is 
the same for all the wavelet layers ($1/128$~sec $\times \; 64$~Hz).
Therefore, the wavelet series $W_{ij}$ can be displayed as a time-frequency
scalogram consisting of 64 wavelet layers with $n=15360$ pixels (data samples) 
each.
This tiling is 
different from the one in the conventional dyadic wavelet
decomposition where the time resolution adjusts to the scale (frequency)
~\cite{ref:wbmethod,ingrid,chatterjiCQG}.
The constant time-frequency resolution makes the 
WaveBurst scalograms similar to spectrograms produced with windowed Fourier 
transformations.

For each layer we first select a fixed fraction $P$ of pixels with the
largest absolute amplitudes. These are called {\it black pixels}.
The number of selected black pixels is
$nP$. All other wavelet pixels are called {\it white pixels}. 
Then we calculate rank statistics for the black pixels within each layer. 
The rank $R_{ij}$ is an integer number from 1 to $nP$, with the rank 1
assigned to the pixel with the largest absolute amplitude in the layer. 
Given the rank of wavelet amplitudes $R_{ij}$, the following non-parametric
pixel statistic is computed
\begin{equation}\label{eq:rsig}
   y_{ij} = -\ln \left( \frac{R_{ij}}{nP} \right).
\end{equation}
For white pixels the value of $y_{ij}$ is set to zero.
The statistic $y_{ij}$ can be interpreted as the pixel's
{\it logarithmic significance}. 
Assuming Gaussian detector noise, the logarithmic significance can be also
calculated as
\begin{equation}\label{eq:pL}
   \tilde{y}_{ij} = g_{P}(\tilde{w}_{ij}) \equiv \ln(P) - \ln \left( 
     \sqrt{2/\pi} \int_{\tilde{w}_{ij}}^{\infty}  
     e^{-x^2/2} \; dx \right) ,
\end{equation}
where $\tilde{w}_{ij}$ is the absolute value of the pixel amplitude in units of
the noise standard deviation. In practice, the LIGO detector noise is not
Gaussian and 
its probability  distribution function is not determined
{\it a priori}.
Therefore, we use the
non-parametric statistic $y_{ij}$, which is a more robust measure of the pixel
significance than $\tilde{y}_{ij}$. 
Using the inverse function of $g_P$ with $y_{ij}$ as an argument,
we introduce the {\it non-parametric amplitude}
\begin{equation}\label{eq:amp}
   w_{ij} = g_P^{-1}(y_{ij}),
\end{equation}
and the {\it excess power ratio} 
\begin{equation}\label{eq:ratio}
   \rho_{ij} = w^2_{ij}-1,
\end{equation}
which characterizes the pixel excess power above the average detector noise.

After the black pixels are selected, 
we require their time-coincidence in the two channels. 
Given a black pixel of significance $y_{ij}$ in the first channel,
this is accepted if the significance of neighboring
(in time) pixels in 
the second channel ($y'_{ij}$)  satisfies
\begin{equation}
   y'_{(i-1)j}+y'_{ij}+y'_{(i+1)j} > \eta,
\end{equation}
where $\eta$ is the {\it coincidence threshold}. Otherwise, the pixel 
is rejected. This  procedure is repeated for all the black pixels in the
first channel. 
The same coincidence algorithm is applied to pixels in the second channel.
As a result, a considerable number of black pixels in both channels
produced by fluctuations of the detector noise are rejected. 
At the same time, black pixels produced by gravitational wave bursts have a high
acceptance probability because of the coherent excess of power in two
detectors. 

After the coincidence procedure is applied to both channels a clustering
algorithm is applied jointly to the two channel pixel maps.
As a first step, we merge (OR) the black 
pixels from both channels into one time-frequency plane.
For each black pixel we define neighbors (either black or white), which
share a side or a vertex with the black pixel. The white neighbors are
called {\it halo} pixels. We define a cluster as a group of black and halo
pixels which are connected either by a side or a vertex.
After the cluster reconstruction, we go back to the original time-frequency
planes and calculate the cluster parameters separately for each channel.
Therefore, there are always two clusters, one per channel, which form a
WaveBurst trigger.

The cluster parameters are calculated using black pixels only. 
For example, the cluster size $k$ is defined as the number of black 
pixels. Other parameters which characterize the cluster 
strength are the cluster excess power ratio
$\rho$ and the cluster {\it logarithmic
likelihood} $Y$.
Given a cluster $C$, these are estimated
by summing over the black pixels in the cluster:    
\begin{equation}
   \rho = \sum_{ij\in{C}} \rho_{ij}~~ , ~~Y = \sum_{ij\in{C}} y_{ij}.
\end{equation}
Given the times $t_i$ of individual pixels, the cluster center time is
calculated as 
\begin{equation}
   T = {\sum_{ij\in{C}} {t_i \; w^2_{ij}}} \; / {\sum_{ij\in{C}} {w^2_{ij}}}.
\end{equation}
As configured for this analysis, WaveBurst initially generated
triggers with frequency content between 64~Hz and 4096~Hz.
As we will see below, the cluster size, likelihood, and excess power
ratio can be used for the further selection of triggers, while the
cluster time and frequency span are used in a coincidence requirement.
The frequency band of interest for this analysis, 100--1100~Hz, is
selected during the later stages of the analysis.

There are two main WaveBurst tunable input parameters: 
the black pixel fraction $P$ 
which is applied to each frequency layer,
and the coincidence threshold $\eta$.
The purpose of these parameters is to control the average black pixel 
occupancy $O(P,\eta)$, 
the fraction of black pixels over the entire time-frequency scalogram.
To ensure robust cluster reconstruction, the occupancy should not be greater
than $1$\%.
For white Gaussian detector noise the functional form of $O(P,\eta)$ can be 
calculated analytically.
This can be used to set a constraint on $P$ and $\eta$
for a given target $O(P,\eta)$.
If $P$ is set too small (less then a few percent), noise
outliers due to instrumental glitches may 
monopolize the limited number of available black pixels and thus
allow gravitational wave signals to remain hidden.
To avoid this domination of instrumental glitches,
we run the analysis with $P$ equal to 10\%.
This value of $P$
together with the occupancy target $O(P,\eta)$ of 0.7\% defines the
coincidence threshold $\eta$ at 1.5.

All the tuning of the WaveBurst method was performed on the S2
playground data set (Section~\ref{sec:dataselect}).
For the selected values of $P$ and $\eta$, the average 
trigger rate per LIGO instrument pair was approximately $6$~Hz,
about twice the false alarm
rate expected for white Gaussian detector noise. 
The trigger rate was further reduced by imposing cuts on the
excess power ratio $\rho$.
For clusters of size $k$ greater than $1$ we required
$\rho$ to be greater than $6.25$ while for
single pixel clusters ($k=1$) we used a more restrictive cut
of $\rho$ greater than $9$.
This selection on the event parameters further reduced the
counting rates per LIGO instrument pair to $\sim{1}$~Hz.
The times and reconstructed parameters of WaveBurst events passing
these criteria were written onto disk.
This allowed the further processing and selection of these events
without the need to re-analyze the full data stream, a process
which is generally time and CPU intensive.

\subsection{Triple coincidence}
\label{ss:triplecoinc}

Further selection of WaveBurst events proceeds by identifying
triple coincidences.
The output of the WaveBurst ETG is a set of coincident triggers for a selected 
interferometer pair $A,B$. Each WaveBurst trigger consists of two clusters,
one in $A$ and one in $B$.
For the three LIGO interferometers there are three possible pairs:
(L1,H1), (H1,H2) and (H2,L1).
In order to establish triple coincidence events, we require 
a time-frequency coincidence of the WaveBurst triggers generated for these
three pairs.
To evaluate the time coincidence we first construct $T_{AB}=(T_{A}+T_{B})/2$,
\ie, the average central time of the $A$ and $B$ clusters
for the trigger.
Three such combined central times are thus constructed: $T_{L1H1}$, $T_{H1H2}$,
and $T_{H2L1}$.
We then require that all possible differences of these combined central times
fall within a time window $T_w=20$~ms.
This window is large enough to accommodate the 
maximum difference in gravitational wave arrival times at the two
detector sites ($10$~ms) and the intrinsic
time resolution of the WaveBurst algorithm which has
an rms on the order of $3$~ms as discussed in
section~\ref{sec:efficiency}. 

We apply a loose requirement on the frequency consistency of the
WaveBurst triggers.
First, we calculate the minimum ($f_{min}$) and maximum ($f_{max}$)
frequency for each interferometer pair $(A,B)$
\begin{equation}
	f_{min}=min(f_{low}^A,f_{low}^B),~~~f_{max}=max(f_{high}^A,f_{high}^B),
\end{equation}
where $f_{low}$ and $f_{high}$ are the low and high frequency boundaries of
the $A$ and $B$ clusters.
Then, the trigger frequency bands are calculated as $f_{max}-f_{min}$ for all
pairs.
For the frequency coincidence, the bands of all three WaveBurst triggers are
required to overlap.
An average frequency is then calculated from the clusters,
weighted by signal-to-noise ratio, and
the coincident event candidate is kept for this analysis if this
average frequency is above 64~Hz and below 1100~Hz.

The final step in the coincidence analysis of the WaveBurst events involves
the construction of a single measure of their combined significance.
As we described already, triple coincidence events consist of three WaveBurst
triggers involving a total of six clusters.
Each cluster has its parameters calculated on a per-interferometer basis.
Assuming white detector noise, the variable $Y$ for
a cluster of size $k$ follows a Gamma probability
distribution.
This motivates the use of the following
measure of the {\it cluster significance}:
\begin{equation} \label{eq:clustersig}
   Z = 
   Y - \ln{ \left( \sum_{m=0}^{k-1}\frac{Y^{m}}{m!} \right), }
\end{equation}
which is 
derived from the logarithmic likelihood $Y$ of a cluster $C$ and from the
number $k$ 
of black pixels in that cluster~\cite{ref:wbmethod2,ref:wbmethod}. 
Given the significance of the six clusters, we compute the
{\it combined significance} of the triple coincidence event as
\begin{equation} \label{eq:combsig}
   Z_G = \left( Z^{L1}_{L1H1} \; Z^{H1}_{L1H1} \; Z^{H2}_{H2L1} \; 
                Z^{L1}_{H2L1} \; Z^{H1}_{H1H2}~\; Z^{H2}_{H1H2} 
         \right)^{1/6},
\end{equation}
where $Z^{A}_{AB}$ ($Z^{B}_{AB}$) is the significance of
the $A$ ($B$) cluster 
for the $(A,B)$ interferometer pair.

\begin{figure}[!thb]
\includegraphics[width=0.98\linewidth]{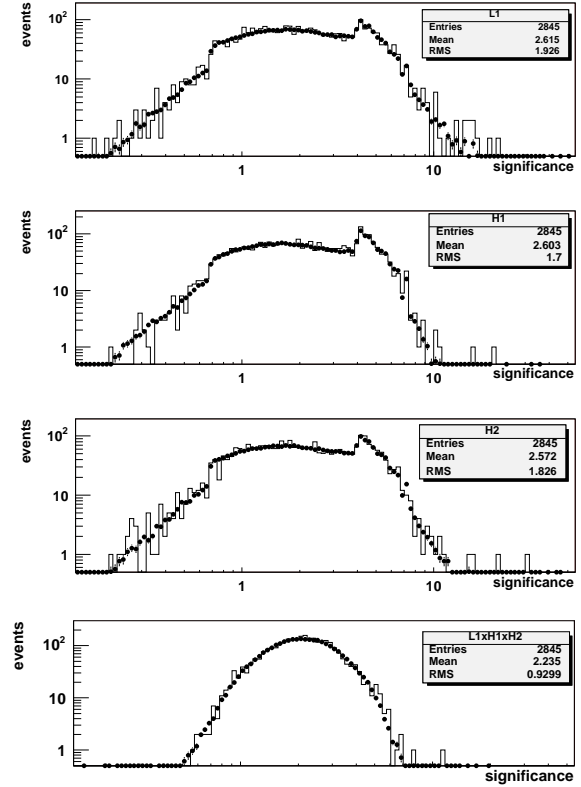}
\caption{The significance distribution of the triple coincident WaveBurst
  events for individual detectors (L1, H1, H2) and the combined significance of
  their triple coincidences (L1xH1xH2) for the S2 data set.
  Solid histograms reflect the zero-lag events, while the points
  represent 
  background (time-lag) events as produced with unphysical time shifts
  between the
  Livingston and Hanford detectors (and normalized to the S2 live-time).
  The change in the significance distribution for the individual
  detectors around significance equal to four is attributed to the onset
  of single pixel clusters (for which a higher threshold was applied).
}
\label{fig:signif}
\end{figure}

In order to evaluate the rate of accidental
coincidences, we have repeated the above analysis
on the data after introducing an unphysical time shift (``lag'')
in the Livingston data stream relative to the Hanford data streams.
The Hanford data streams are not shifted relative to one another, so
any noise correlations from the local environment are preserved.
Figure~\ref{fig:signif} shows the distribution of cluster
significance (equation~\ref{eq:clustersig}) from the three individual
detectors, and the combined significance
(equation~\ref{eq:combsig}),
over the entire S2 data set, for both zero-lag
and time-lag coincidences.
Using 46 such time-lag instances of the
S2 playground data we have set the
threshold on $Z_G$ for this search in order to yield a targeted
false alarm rate of $10~\mu$Hz.
Without significantly compromising the pipeline sensitivity, this
threshold was selected to be $\ln(Z_G)>1.7$.
In the 64--1100~Hz frequency band,
the resulting false alarm rate in the S2 playground analysis
was approximately $15~\mu$Hz.
The coincident events selected by WaveBurst in this
way are then checked
for their waveform consistency using the $r$-statistic.

\subsection{$r$-statistic test}\label{sec:rstat}

The $r$-statistic test~\cite{rstatCQG} is applied as the
final step of searching for gravitational wave event candidates.
This test re-analyzes the raw (unprocessed) interferometer data
around the times of coincident events identified by the
WaveBurst ETG.

The fundamental building block in performing this waveform consistency
test is the $r$-statistic, or the normalized linear correlation 
coefficient of two sequences, $\{ x_i \}$ and  $\{ y_i \}$
(in this case, the two gravitational wave signal time series):
\begin{equation}
\label{eqn:rstat}
r = \frac {\sum_i (x_i - \bar{x})(y_{i} - \bar{y})} 
{\sqrt{\sum_i (x_i-\bar{x})^2}\sqrt{\sum_i (y_{i}-\bar{y})^2}},
\end{equation}
where $\bar{x}$ and $\bar{y}$ are their respective mean values.
This quantity assumes values between $-1$ for fully anti-correlated
sequences and $+1$ for fully correlated sequences.
For uncorrelated white noise, we expect 
the $r$-statistic values obtained for arbitrary sets of points of
length $N$
to follow a normal distribution with zero mean 
and $\sigma=1/\sqrt{N}$.
Any coherent component in the two sequences will cause $r$ to deviate from
the above normal distribution.
As a normalized quantity, the $r$-statistic
does not attempt to measure the consistency between the relative amplitudes
of the two sequences.
Consequently, it offers the advantage of
being robust against
fluctuations of detector amplitude response and noise floor. 
A similar method based on this type of time-domain cross-correlation
has been implemented in a LIGO search for
gravitational waves associated with a GRB~\cite{exttrigCQG,ref:GRB030329}
and elsewhere~\cite{explnaut}.

As will be described below, the final output of the $r$-statistic test
is a combined confidence statistic which is constructed from
$r$-statistic values calculated for all three pairs of
interferometers.  For each pair, we use only the absolute value of the
statistic, $|r|$, rather than the signed value.  This is because
an astrophysical signal can produce either a correlation or an
anticorrelation in the interferometers at the two LIGO sites,
depending on its sky position and polarization.
In fact, the $r$-statistic analysis was done using whitened (see below)
but otherwise uncalibrated data, with an arbitrary sign convention.
A signed correlation test using calibrated data would be appropriate for the
H1-H2 pair, but all three pairs were treated equivalently in the
present analysis.

The number of points $N$ considered in calculating the statistic
in Eq.~(\ref{eqn:rstat}), or equivalently the \emph{integration time} $\tau$, 
is the most important parameter in the construction of the 
$r$-statistic.
Its optimal value depends in general on the duration of the
signal being considered for detection.
If $\tau$ is too long, the candidate signal is ``washed out'' by
the noise when computing $r$.
On the other hand, if it is too short, then only part of the coherent
signal is included in the integration.
Simulation studies have shown that most of the short-lived signals
of interest to the LIGO burst search can be identified successfully
using a set of three discrete integration times with lengths of
20, 50 and 100~ms.

Within its LIGO implementation, the $r$-statistic analysis first
performs data ``conditioning'' to restrict the frequency content of
the data to LIGO's most sensitive band and to 
suppress any coherent
lines and instrumental artifacts.
Each data stream is first band-pass filtered with an 8th-order
Butterworth filter with corner frequencies of 100~Hz and 1572~Hz, then
down-sampled to a 4096~Hz sampling rate.  The upper frequency of
1572~Hz was chosen in order to have 20~dB suppression at 2048~Hz and
thus avoid aliasing.  The lower frequency of 100~Hz was chosen to
suppress the contribution of seismic noise; it also defines the lower
edge of the frequency band for this gravitational wave burst search,
since it is above the lower frequency limit of 64~Hz for WaveBurst
triggers.
The band-passed data are then whitened with a linear
predictor error filter with a 10~Hz resolution
trained on a 10~second period before the event start time.
The filter removes predictable content, including lines that were
stationary over a 10~second time scale. It also has the effect of
suppressing frequency bands with large stationary noise, thus emphasizing
transients~\cite{chatterjiCQG}. 

The next step in the $r$-statistic analysis involves the
construction of all
the possible $r$ coefficients given the number of interferometer
pairs involved in the trigger, their possible relative time-delays due
to their geographic separation, and the various integration times
being considered.
Relative time delays up to $\pm10$~ms are considered for each detector
pair, corresponding to the light travel time between the Hanford and
Livingston sites. Future analyses will restrict the time delay to a
much smaller value when correlating data from the two Hanford
interferometers, to allow only for time calibration uncertainties.
Furthermore,
in the case of WaveBurst triggers with reported durations greater than
the integration time $\tau$, multiple integration {\em windows} of
that length are considered, offset from the reported start time of the
trigger by multiples of $\tau/2$.  For a given integration window
indexed by $p$ (containing $N_p$ data samples), ordered pair of
instruments indexed by $l,m$ $(l\neq m)$, and relative time delay
indexed by $k$, the $r$-statistic value $|r^k_{plm}|$ is calculated.
For each $p\,l\,m$ combination, the {\em distribution} of
$|r^k_{plm}|$ for all values of $k$ is compared
to the null hypothesis
expectation of a normal distribution with zero mean and
$\sigma=1/\sqrt{N_p}$ using the Kolmogorov-Smirnov test.  If these are
statistically consistent at the 95\% level, then the algorithm assigns
no significance to any apparent correlation in this detector pair.
Otherwise, a one-sided significance and its associated logarithmic
confidence are calculated from the {\em maximum} value of
$|r^k_{plm}|$ for any time delay, compared to what would be expected
if there were no correlation.  Confidence values for all ordered
detector pairs are then averaged to define the combined correlation
confidence for a given integration window.
The final result of the $r$-statistic test, $\Gamma$, is the maximum
of the combined correlation confidence over all of the integration
windows being considered.
Events with a value of $\Gamma$ above a given threshold
are finally selected.

The $r$-statistic implementation, filter parameters, and set of
integration times were chosen based on their performance for various
simulated signals.  The single remaining parameter, the threshold on
$\Gamma$, was tuned primarily in order to ensure that much less than
one background event was expected in the whole S2 run, corresponding
to a rate of O$(0.1)~\mu$Hz.  Since the rate of WaveBurst triggers was
approximately $15~\mu$Hz, as mentioned in
Section~\ref{ss:triplecoinc}, a rejection factor of around 150 was
required.

Table~\ref{t:reject} shows the rejection efficiency of the
$r$-statistic test for two thresholds on $\Gamma$ when the test is
applied to white Gaussian noise (200 ms segments), to real S2
interferometer noise at randomly selected times (200 ms segments), and
to the data at the times of time-lag (\ie, background)
WaveBurst triggers in the S2
playground.  In the first two cases, 200~ms of data was processed by
the $r$-statistic algorithm, whereas in the latter case, the amount of
data processed was determined by the trigger duration reported by
WaveBurst.  The table shows that random detector noise rarely produced
a $\Gamma$ value above $3.0$, but the rejection factor for WaveBurst
triggers was not high enough.  A $\Gamma$ threshold of $4.0$ was
ultimately chosen for this analysis, yielding an estimated rejection
factor of $\sim 250$ for WaveBurst triggers.
\begin{table} [htb]
\begin{center}
\caption[]{Percentage of S2 background events rejected by 
the $r$-statistic for two different thresholds on $\Gamma$.
}
\label{t:reject}
\small
\begin{tabular}{lll}
\hline\hline
Event Production & $\Gamma>3.0$ & $\Gamma>4.0$ \\
\hline
200 ms white Gaussian noise     & $99.9992$\% & $99.999996$\%   \\
200 ms real noise (random)      & $99.89$\%   & $99.996$\%   \\
WaveBurst  background events    & $98.6 \pm 0.5$\%   & $99.6 \pm 0.3$\%       \\
\hline\hline
\end{tabular}
\end{center}
\end{table} 
As we will discuss in Section~\ref{sec:efficiency},
the $r$-statistic waveform consistency test with $\Gamma>4.0$
represents, for the waveforms we considered, a sensitivity that is equal to 
or better than that of the WaveBurst ETG.
As a result of this, the false dismissal probability 
of the $r$-statistic test does not impair the efficiency of the whole pipeline.

\section{Vetoes}\label{sec:vetoes}

We performed several studies in order to establish any correlation
of the triggers produced by the WaveBurst search algorithm
with environmental and instrumental glitches.
LIGO records hundreds of auxiliary read-back channels of the
servo control systems employed in the instruments' interferometric
operation as well as auxiliary channels monitoring the instruments'
physical environment.
These channels can provide ways
for establishing evidence that a transient is not of astrophysical origin, \ie,
a glitch attributed to the instruments themselves and/or to their
environment.
Assuming that the coupling of these channels to a genuine gravitational
wave burst is null (or below threshold within the context
of a given analysis), such glitches appearing in these auxiliary
channels may be used to veto the events that appear simultaneously
in the gravitational wave channel.

Given the number of auxiliary channels and the parameter space that we
need to explore for their analysis, an exhaustive {\it{a priori}}
examination of all of them is a formidable task.
The veto study was limited to the S2 playground data set and to a
few tens of channels thought to be most relevant.
Several different choices of filter and threshold parameters
were tested in running the glitch finding algorithms.
For each of these configurations, the efficiency of the auxiliary 
channel in vetoing the event triggers (presumed to be glitches), as well as
the dead-time introduced by using that auxiliary channel as a veto,
were computed and compared to judge the effectiveness of the veto condition.

Another important consideration in a veto analysis is to verify the
absence of coupling between a real gravitational wave burst and the
auxiliary channel, such that the real burst could cause itself to be
vetoed.  The ``safety'' (absence of such a coupling) of veto
conditions was evaluated using hardware signal injections (described
in section~\ref{sec:efficiency}), by checking whether the simulated
burst signal imposed on the arm length appeared in the auxiliary
channel.  Only one channel, referred to as AS\_I, in the L1 instrument
derived from the antisymmetric port
photodiode with a demodulation phase orthogonal to that of the
gravitational wave channel, was found to be ``unsafe'' in this
respect, containing a small amount of the injected signal.
 
None of the channels and parameters we examined yielded an obviously
good veto (\eg, one with an efficiency of 20\% or greater and a
dead-time of no more than a few percent) to be used in this search.
Among the most interesting channels was
the one in the L1 instrument that recorded the DC level of the light out
of the antisymmetric port of the interferometer, referred to as AS\_DC.
That channel was seen to correlate with the gravitational
wave channel through a non-linear coupling with interferometer 
alignment fluctuations.
A candidate veto based on this channel was shown to be able to reject
$\sim 15$\% of the triggers, but with a non-negligible dead-time of 5\%.
Finding no better option, we decided not to apply any {\it a priori}
vetoes in this search,
judging that the effect on the results would be insignificant.

Although none of the auxiliary channels studied in the playground
data yielded a compelling veto, these studies provided experience
applicable to examining
any candidate gravitational wave event(s) found in the full data set.
A basic principle established for the search was that a
statistical excess of zero-lag event candidates (over the expected background)
would not, by itself, constitute a detection; the candidate(s) would
be subjected to further scrutiny to rule out any environmental or
instrumental explanation that might not have been apparent in the
initial veto studies.
As will be described in the next section, one event did survive
all the pre-determined cuts of the analysis but
subsequent examination of auxiliary channels identified an
environmental origin for the signal in the two Hanford detectors.
\section{Signal and Background Rates}\label{sec:evana}

In the preceding section we described the methods
that we used for the selection of burst events.
These were applied to the
S2 triple coincidence data set excluding the playground
for a total of
$239.5$~hours ($9.98$~days) of observation time.
Every
aspect of the analysis discussed from this point on
will refer only to this data set.

\subsection{Event analysis}

The WaveBurst analysis applied to the S2 data yielded 16
coincidence events (at zero-lag).
The application of the $r$-statistic cut rejected 15 of
them, leaving us with a single event that passed all the
analysis criteria.

\begin{figure}[!thb]
\includegraphics[width=1.00\linewidth]{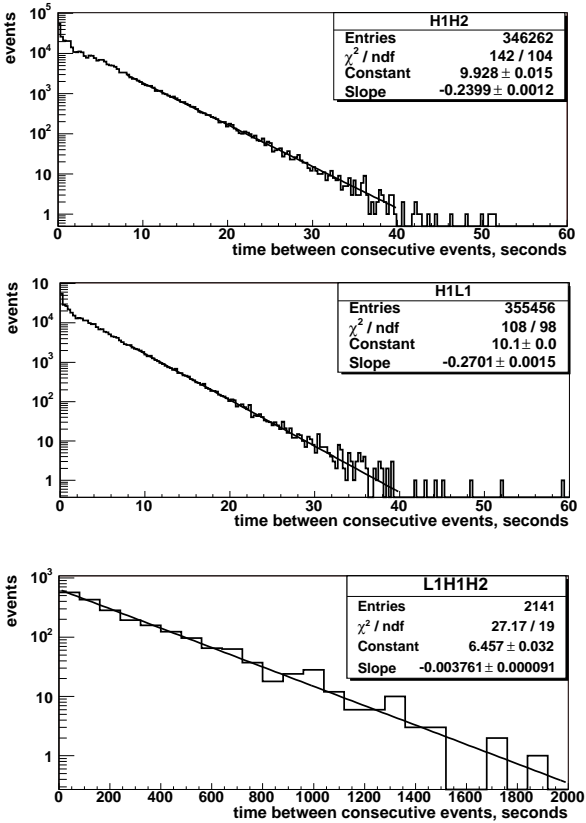}
\caption{
Time between consecutive WaveBurst events (prior to the application
of the $r$-statistic test).  The top two panels show the distributions
for double-coincidence H1-H2 and H1-L1 triggers, respectively.
The triple coincidence events, shown in the bottom panel, are
reasonably well described by a Poisson process of constant mean.
The exponential fits are performed for time delays greater than
4~s.
}
\label{fig:tbe}
\end{figure}

The background in this search
is assumed to be due to random coincidences
between unrelated triggers at the two LIGO sites.   
We have measured this background by artificially shifting the raw
time series of the L1 instrument.
As in our S1 search, we have chosen not to time-shift relative
to each other the two Hanford instruments (H1, H2).
Although we had no evidence of H1-H2 correlations in the S1 burst
search, indications for such correlations in other LIGO searches
exist~\cite{ref:SUL}.
A total of 46 artificial lags of the raw time series of the L1 instrument,
at 5-second steps in the range [$-115$,$115$] seconds,
were used in order to make a measurement of the accidental rate of
coincidences, \ie, the background.
This step size was much larger than the duration of any signal that we
searched for and was also larger than the autocorrelation time-scale for
the trigger generation algorithm applied to S2 data.
This can be seen in Fig.~\ref{fig:tbe} where 
a histogram of the time between
consecutive events is shown for the double and their
resulting triple coincidence WaveBurst zero-lag events before any combined
significance or $r$-statistic cut is applied.
These distributions follow the expected exponential form, indicating
a quasi-stationary Poisson process.
The background events generated in this way were also subjected
to the $r$-statistic test in an identical way with the one used
for the zero-lag events.
Each time-shift experiment had a different live-time according to the
overlap, when shifted, of the many non-contiguous data segments that
were analyzed for each interferometer.  Taking this into account,
the total effective live-time for the purpose of measuring
the background in this search was $391$~days, equal to $39.2$ times
the zero-lag observation time.

\begin{figure}[!thb]
\includegraphics[width=1.00\linewidth]{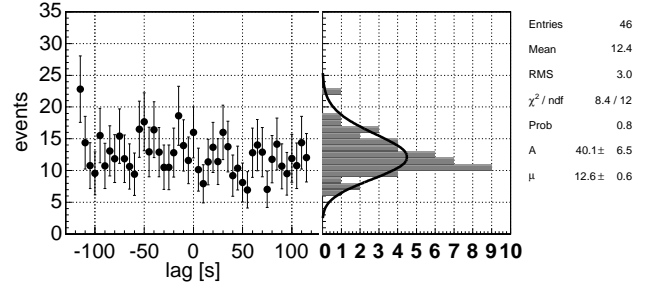}
\caption{
WaveBurst event count (prior to the $r$-statistic test)
versus lag time (in seconds) of the L1
interferometer with respect to H1 and H2.
The zero-lag
measurement, \ie, the only coincidence measurement that is physical,
is also shown.
Due to the fragmentation of the data set, each
time-lag has a slightly different live-time; for this reason,
the event count is corrected so that they all correspond
to the zero-lag live-time.
A projection of the event counts to a one-dimensional histogram
with a Poisson fit is also shown in the adjacent panel.
}
\label{fig:background}
\end{figure}

A plot of the measured background events found in each of the
46 time-lag experiments, {\em before} the application of the $r$-statistic,
is shown in Fig.~\ref{fig:background} as a function of the lag time.
These numbers of events are corrected so that they all correspond to the 
zero-lag live-time.
A Poisson fit can be seen in the adjacent
panel; the fit describes the distribution of event counts reasonably well.

\begin{figure}[!thb]
\includegraphics[width=1.00\linewidth]{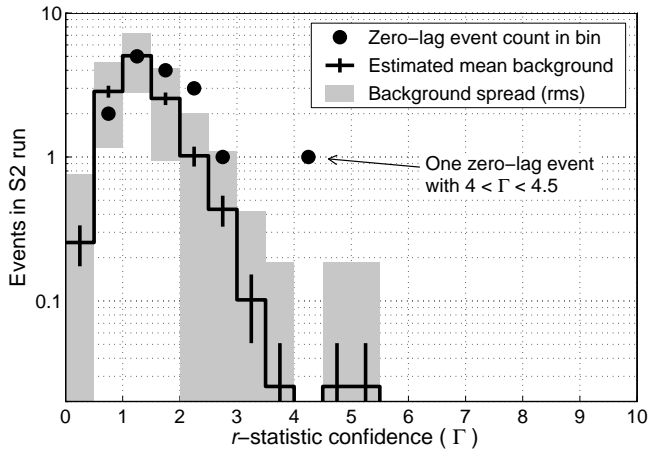}
\caption{
Circles: histogram of $r$-statistic confidence value ($\Gamma$) for
zero-lag events passing the WaveBurst analysis.
Stair-step curve: mean background per bin, estimated from time lags,
for an observation time equal to that of the zero-lag analysis.
The black error bars indicate the statistical uncertainty on the mean
background.
The shaded bars represent the expected root-mean-square statistical
fluctuations on the number of background events in each bin.
}
\label{fig:background_rstat}
\end{figure}

Figure~\ref{fig:background_rstat} shows a histogram
of the $\Gamma$ values, \ie,
the multi-interferometer combined correlation confidence,
for the zero-lag events.
The normalized background distribution, estimated from time-lag
coincidences, is shown for comparison.  One zero-lag event passed the
requirement $\Gamma>4$ that we had chosen based on the playground
data; this event will be discussed in the following
subsection.  Only two time-lag coincidences above this $\Gamma$
threshold were found among all 46 time lags.  With such low statistics,
the rate and distribution of the background for large $\Gamma$ is
poorly known, but we can get an approximate measure of the
significance of the zero-lag event by comparing it to the cumulative
mean background rate with $\Gamma>4$, which is roughly $0.05$~events
for the same observation time.  Thus, the chance of having found such 
a background event in the zero-lag sample is roughly 5\%.
Table~\ref{tbl:eventresults} summarizes the number of events and
corresponding rates before and after the application of the
$r$-statistic.  The background estimates reported in the table are
normalized to the same live-time as for the zero-lag coincidence
measurement.

\begin{table}[!thb]
\caption{Event statistics for the S2 burst search.
The expected numbers of background events are normalized to the
live-time of the zero-lag analysis.}
\label{tbl:eventresults}
\begin{tabular}{lcc}
\hline\hline
   WaveBurst & events in $239.5$~hours ($9.98$~days) & rate \\
\hline
 & Before $r$-statistic test & \\
Coincidences & 16    & $18.6 \; \mu$Hz\\
Background   & 12.3  & $14.3 \pm 0.7\; \mu$Hz\\
 & After $r$-statistic test& \\
Coincidences & 1     & $1.2 \; \mu$Hz\\
Background   & 0.05  & $0.06 \pm 0.04\; \mu$Hz\\
\hline\hline
\end{tabular}
\end{table}

Sources of systematic errors may arise in the choices we have made on
how to perform the time-lag experiments, namely the choice of step
and window size as well as the time-lag method by itself.
We have performed time-lag experiments using different
time steps, all of which yielded statistically consistent results.
The one-sigma systematic uncertainty from the choice of step size
is estimated to be less than 0.04 events with $\Gamma>4$.
 
\subsection{Examination of the surviving event candidate}

The single event in the triple coincidence data set
that survived all previously described analysis cuts
barely passed the WaveBurst combined significance
and $r$-statistic thresholds.
An examination of the event parameters estimated by
WaveBurst revealed that the three instruments recorded
low frequency signals in the $\sim 135$~Hz range and with
comparable bandwidths, although WaveBurst provides only a rough
estimate of the dominant frequency of an event candidate.
The signal $\hrss$ strengths  in the two Hanford detectors
were in the $6 \times 10^{-20}-10^{-19}$~\hrssu\ range, 
well above
the instruments' typical noise in this band, while
for the Livingston detector, $\hrss$ was at the
$2.7 \times 10^{-21}$~\hrssu\
level, much closer to the noise floor of the instrument.

Given the low estimated probability of this event being due to a
random triple coincidence, it was treated as a candidate
gravitational wave detection and was therefore subjected to additional
scrutiny.  In particular, the auxiliary interferometric and environmental
monitoring channels were examined around the time of the event to
check for an interferometer malfunction or an environmental cause.
The investigation revealed that the event occurred during a
period of strongly elevated acoustic noise at Hanford lasting tens of
seconds, as measured by microphones placed near the interferometers.
The effects of environmental influences on 
the interferometers were measured in a special study during the
S2 run by intentionally generating acoustic
and other environmental disturbances and comparing the resulting
signals in the gravitational wave and environmental monitoring
channels. These coupling measurements 
indicated that the acoustic event recorded on the microphones
could account for the amplitude and frequency of the signal in
the H1 and H2 gravitational wave channels at the time of the 
candidate event.
On this basis, it was clear that the candidate event should be
attributed to the acoustic disturbance and not to a gravitational wave.

The source of the acoustic noise appears to have been an aircraft.
Microphone signals from the five Hanford buildings exhibited Doppler
frequency shifts in a sequence consistent with the overflight of an
airplane roughly paralleling the X arm of the
interferometers, on a typical approach to the nearby Pasco, Washington
airport. Similar signals in the microphone and gravitational wave
channels at other times have been visually confirmed as over-flying
airplanes.

No instrumental or environmental cause was identified for the signal
in the Livingston interferometer at the time of the candidate event,
but that signal was much smaller in amplitude and was consistent with
being a typical fluctuation in the Livingston detector noise,
accidentally coincident with the stronger signals in the two Hanford
detectors.

Because of the sensitivity of the interferometers to the acoustic
environment during S2, a program to reduce acoustic coupling was
undertaken prior to S3. The acoustic sensitivities of the
interferometers were reduced by 2 to 3 orders of magnitude by
addressing the coupling mechanisms on optics tables located
outside of the vacuum system, and by acoustically isolating the
main coupling sites.

\subsection{Propeller-airplane acoustic veto}

\begin{figure}[!thb]
\includegraphics[width=1.00\linewidth]{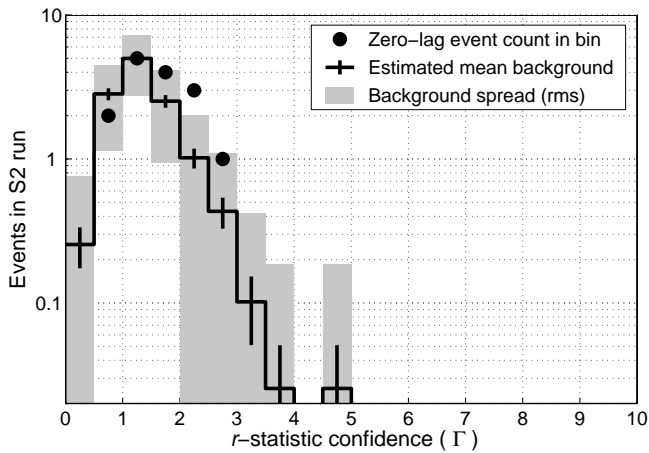}
\caption{
Circles: histogram of $r$-statistic confidence value ($\Gamma$) for
zero-lag events passing the WaveBurst analysis and surviving the
acoustic veto.
The background estimated from time-lag events surviving the acoustic
veto is shown in the same manner as in Fig.~\ref{fig:background_rstat}.
}
\label{fig:background_rstat_veto}
\end{figure}

Given the clear association of the surviving event with an acoustic
disturbance, we tracked the power in a particular microphone channel,
located in the LIGO Hanford corner station, over the entire S2 run.
We defined a set of time intervals with significantly elevated
acoustic noise by setting a threshold on the power in the 62--100~Hz
band---where propeller airplanes are observed to show up most
clearly---averaged over one-minute intervals.  The threshold was chosen by
looking at the distribution over the entire S2 run, and was far below
the power at the time of the ``airplane'' outlier event discussed
above.  Over the span of the run, $0.7$\% of the data was collected
during times of elevated acoustic noise as defined in this way.
Eliminating these time intervals removes the zero-lag outlier as well
as the time-lag event with the largest value of $\Gamma$, while having
only a slight effect on the rest of the background distribution, as
shown in Fig.~\ref{fig:background_rstat_veto}.
We conclude that acoustic disturbances from propeller airplanes
contribute a small but non-negligible background if this veto is not
applied.

\subsection{Rate limit}
\label{ss:ratelimit}

We now use the results of this analysis to
place a limit on the average rate (assuming a uniform
distribution over time) of gravitational wave bursts that are
strong enough to be detected reliably by our analysis pipeline.  The
case of somewhat weaker signals, which are detectable with efficiency
less than unity, will be considered in the next section.

Our intention at the outset of this analysis was to calculate a
frequentist 90\% confidence interval from the observation time, number
of observed candidate events, and estimated background using the
Feldman-Cousins~\cite{feldman98a} approach.  Although this procedure
could yield an interval with a lower bound greater than zero, we would
not claim the detection of a gravitational wave signal based on that
criterion alone; we would require a higher level of statistical
significance, including additional consistency tests.  Thus, in the
absence of a detection, our focus is on the upper bound of the
calculated confidence interval; we take this as an upper limit
on the event rate.

The actual outcome of our analysis presented us with a
dilemma regarding the calculation of a rate limit.
Our pipeline was designed to perform a ``blind'' upper limit analysis, with all
choices about the analysis to be based on playground data which
was excluded from the final result; following this principle, the
``emergent''
acoustic veto described above should be disallowed (since it was
developed in response to the candidate event which passed all of the
initial cuts), and the upper bound should be calculated based on a
sample of one candidate event.  On the other hand, it seemed
unacceptable to ignore the
clear association of that event with a strong acoustic disturbance
and to continue to treat it as a candidate gravitational wave burst.
We decided to apply the acoustic veto, reducing the
observation time by $0.7$\% and calculating an upper limit based on a
final sample containing no events.  However, any decision to alter the
analysis procedure based on information from the analysis must be
approached with great caution and an awareness of the impact on the
statistics of the result.  In particular, a frequentist confidence
interval construction which has been
designed to give 90\% minimum coverage for an
ordinary (unconditional) analysis procedure can yield {\em less than}
90\% coverage if it is blindly used in a conditional analysis
involving an emergent veto, due to the chance that a real
gravitational wave burst could be vetoed, and due to the fact that the
background would be mis-estimated.  In the present analysis, we
know that the chance of a gravitational wave burst being eliminated by
the acoustic veto described above is only $0.7$\%; however, we must
consider the possibility that there are other, ``latent'' veto
conditions which are not associated with any events in this
experimental instance but which might be adopted to veto a
gravitational wave burst in case of a chance coincidence.

It is impossible to enumerate all possible latent veto conditions
without an exhaustive examination of auxiliary channels in the full
data set.  Judging from our experience with examining individual event
candidates and potential veto conditions in the playground data set,
we believe that there are few possible veto conditions with
sufficiently low dead-time and a plausible coupling mechanism (like
the acoustic veto) to be considered.  Nevertheless, we have performed
Monte Carlo simulations to calculate frequentist coverage for various
conditional limit-setting procedures under the assumption that there
are {\em many} latent vetoes, with a variety of individual dead-times
and with a net combined dead-time of 35\%.  A subset of eight latent
vetoes with individual dead-times less than 5\%, sufficiently low that
we might adopt the veto if it appeared to correlate with a single
gravitational-wave event, had a combined dead-time of 12\%.  Veto
conditions with larger dead-times would be considered only if
they seemed to explain multiple event candidates to a degree
unlikely to occur by chance.

The simulations led us to understand that we can preserve
the desired minimum coverage (\eg, 90\%) by assigning a
somewhat larger interval when an emergent veto has been applied.
This is a means of incorporating the information that an observed
event is {\em probably} due to the environmental disturbance
identified by the veto, without assuming that it is {\em certainly}
due to the environmental disturbance and simply applying the veto.
The resulting upper limit is looser than what would be obtained by
simply applying the veto.  Among a number of possible ways to assign
such an interval, we choose to use the Feldman-Cousins interval
calculation with an input confidence level somewhat greater than our
target coverage and with the background taken to be
zero.
Taking the background to be zero provides some necessary conservatism
since we have not sought vetoes for the time-lag coincidences from
which the background was originally estimated, but this has little
effect on the result since the background rate is low.

According to the simulations, using a confidence level of 92\% in the
Feldman-Cousins upper limit calculation after adopting an emergent
veto is sufficient to ensure an actual minimum coverage of greater
than 90\%, and using a confidence level of 96\% is sufficient to
ensure an actual minimum coverage of greater than 95\%.
The resulting rate limits for strong gravitational wave bursts are
presented in Table~\ref{tbl:limitresults}.  The upper limit at 90\%
confidence, $0.26$ events per day, represents an improvement over the
rate limit from our S1 result~\cite{ref:BUL} by a factor of 6.  As
will be described in the following section, the present analysis also
is sensitive to much weaker bursts than the S1 analysis was.

\begin{table}[!thb]
\caption{
Upper limits on the rate of strong gravitational wave bursts for two
different frequentist confidence levels.  The method used to calculate
these limits is described in the text.}
\label{tbl:limitresults}
\begin{tabular}{cc}
\hline\hline
   Confidence level & Upper limit \\
\hline
90\%    & $0.26$ events/day \\   
95\%    & $0.33$ events/day \\   
\hline\hline
\end{tabular}
\end{table}

\section{Efficiency of the Search}\label{sec:efficiency}
\subsection{Target waveforms and signal generation}

In order to estimate the sensitivity of the burst analysis pipeline,
we studied its
response to simulated signals of various waveform morphologies and strengths.
The simulated signals were prepared in advance, then ``injected'' into
the S2 triple coincidence data set by using software to add them to
the digitized time series that had been recorded by the
detectors~\cite{ref:S2MDC}.  The times of the simulated signals were
chosen pseudo-randomly, uniformly covering the S2 triple coincidence
data set with an average separation of one minute and a minimum
separation of 10 seconds.  The modified data streams were then
re-analyzed using the same analysis pipeline.

Several {\it{ad hoc}} and astrophysically motivated waveforms were
selected for injections:
\begin{itemize}
\item sine-Gaussian waveforms of the form
$h(t+t_0)=h_{0} \sin(2\pi f_{0} t) \exp(-t^2/\tau^2)$, where 
$\tau$ was chosen according to 
$\tau=$Q$/(\sqrt{2}\pi f_{0})$ with Q=8.9, and  $f_{0}$ 
assumed the value of 100, 153, 235, 361, 554, and 849~Hz; 
\item Gaussian waveforms of the form
$h(t+t_0)=h_{0} \exp(-t^2/\tau^2)$ and with $\tau$ equal to
0.1, 0.5, 1.0, 2.5, 4.0 and 6.0~ms;
\item waveforms resulting from numerical simulations of core collapse
supernovae that are available in the literature~\cite{ref:ZM,ref:DFM,ref:OBLW};
\item binary black hole merger waveforms as described
in~\cite{lazarus,astrogravs} and for
total system masses of 10, 30, 50, 70 and 90 solar masses.
\end{itemize}
The sine-Gaussian and Gaussian waveforms were chosen to represent
the two general classes of short-lived gravitational wave bursts
of narrow-band and
broad-band character respectively. The supernovae and binary black
hole merger waveforms were adopted as a more realistic model for
gravitational wave bursts.

In order to ensure self-consistent injections which would accurately
test the coincidence criteria in the pipeline, we took
into account the exact geometry of the individual LIGO detectors
with respect to the impinging gravitational burst wavefront.
A gravitational wave burst is expected to be comprised of two waveforms
$h_{+}(t)$ and $h_{\times}(t)$ which represent its two polarizations,
conventionally defined with respect to the
polarization of the source.
The signal produced on the output of a LIGO detector $h_{det}(t)$ is a linear
combination of these two waveforms,
\begin{equation}\label{eq:hdet}
h_{det}(t)=F_{+} h_{+}(t) + F_{\times} h_{\times}(t),
\end{equation}
where $F_{+}$ and $F_{\times}$ are the antenna pattern
functions~\cite{thorne87a,ref:Saulson:1994}.
The antenna pattern functions
depend on the source location on the sky
(spherical polar angles $\theta$ and $\phi$) and the wave's polarization
angle $\psi$.
The source coordinates $\theta$ and $\phi$ were chosen randomly
so that they would appear uniformly distributed on the sky.
For every source direction the simulated signals were injected
with the appropriate relative time delay corresponding to the geometric
separation of the two LIGO sites.
For the two {\it{ad hoc}} waveform families (sine-Gaussian, Gaussian)
as well as for the supernovae ones,
a linearly polarized wave was assumed with a random polarization angle.
The binary black hole merger waveforms come with two
polarizations~\cite{lazarus} and both were taken into account.

For the supernovae waveforms the {\em inclination} of the source with respect
to the line of sight was taken to be optimal (ninety degrees), so that
the maximum gravitational
wave emission is in the direction of the Earth.
For the binary black hole merger case we used the same $\hrss$ amplitude
in the two polarizations thus corresponding to an inclination of
59.5 degrees.
Of course, a real
population of astrophysical sources would have random inclinations,
and the wave amplitude at the Earth would depend on the inclination as
well as the intrinsic source strength and distance.  Our injection
approach is in keeping with our intent to express the detection
efficiency in terms of the gravitational wave amplitude reaching the
Earth, not in terms of the intrinsic emission by any particular class
of sources (even though some of the waveforms we consider are derived
from astrophysical models).  For a
source producing radiation in only one polarization state, a
change in the inclination simply reduces the amplitude at the Earth by a
multiplicative factor.  However, a source which emits two distinct
polarization components produces a net waveform at the Earth which
depends nontrivially on inclination angle; thus, our fixed-inclination
injections of black hole merger waveforms can only be considered as
discrete examples of such signals, not as representative of a
population.
In any case, the waveforms we use are only approximations to those
expected from real supernovae and black hole mergers.

\subsection{Software injection results}
In order to add the aforementioned waveforms to the raw detector data,
their signals were first digitized at the LIGO sampling frequency of
16384~Hz. Their amplitudes defined in strain dimensionless units were
converted to units of ADC counts using 
the response functions of the detectors
determined from calibration~\cite{ref:S2calibration}.
The resulting time series of ADC($t$) were then added to the raw
detector data and were made available to the analysis pipeline.
In analyzing the injection data, every aspect of the analysis
pipeline that starts with single-interferometer time series ADC($t$)
and ends with a collection of event triggers
was kept identical to the one that was used in the analysis
of the real, interferometric data, including the acoustic veto.
For each of the four waveform families we introduced earlier
in this section, a total of
approximately 3000 signals were injected into the three 
LIGO detectors, uniformly distributed in time
over the entire S2 data set
that was used for setting the rate bound.
As in our S1 signal injection analysis, we
quantify the strength of the injected signals using the
{\it root-sum-square} (rss) amplitude
{\it{at the Earth}}
(\ie, {\it{without}} folding in the antenna pattern of a detector)
defined by 
\begin{equation}\label{eq:hrss}
\hrss \equiv \sqrt{\int (|h_{+}(t)|^2 + |h_{\times}(t)|^2) \, dt} ~ .
\end{equation}
This is a measure of the square root of the signal ``energy'' and
it can be shown that, when divided by the detector spectral noise,
it approximates the signal-to-noise ratio
that is used to quantify the detectability of a signal in optimal filtering.
The quantity $\hrss$ has units of~\hrssu\
and can thus be directly compared to the detector sensitivity curves,
as measured by power spectral densities over long time scales. 
The pixel and cluster strength quantities calculated by the WaveBurst
ETG are monotonic functions of the $\hrss$ of a given signal.
The $\hrss$ amplitudes of the injected signals were chosen randomly from 20
discrete logarithmically-spaced values in order to map out the detection
efficiency as a function of signal strength.

The efficiency of the analysis pipeline is defined as the fraction of
injected events which are successfully detected.
The software injections exercised a
range of signal strengths that allowed us to measure
(in most cases)
the onset of efficiency up to nearly unity.
Efficiency measurements between $0.01$ and $0.99$
were fitted with an asymmetric sigmoid of the form
\begin{equation}
\epsilon(\hrss) = \frac{1}{1 +
\left(\frac{\hrss}{h_\mathrm{mid}}\right)^{\alpha \left( 1+\beta
\tanh(\hrss/h_\mathrm{mid}) \right) }} \, ,
\end{equation}
where 
$h_\mathrm{mid}$ is the $\hrss$ value corresponding to an
efficiency of $0.5$,
$\beta$ is the parameter that describes the asymmetry of the sigmoid
(with range $-1$ to $+1$),
and $\alpha$ describes the slope.
The analytic expressions of the fits were then used to determine the signal
strength $\hrss$ for which an efficiency of 50\%, 90\% and 95\% was
reached.

\begin{figure}[!thb]
\includegraphics[width=0.98\linewidth]{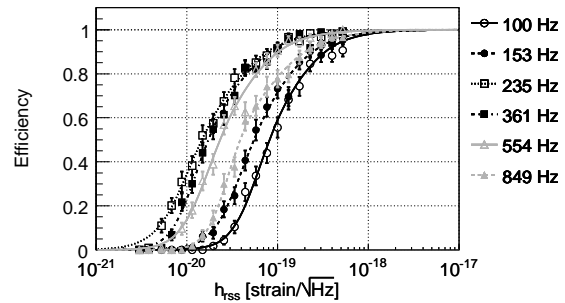}
\caption{Detection efficiency of the analysis pipeline as a function
of the signal strength for sine-Gaussian waveforms of Q=8.9 and
central frequencies of 100, 153, 235, 361, 554 and 849~Hz.
The efficiencies plotted reflect averaging over random sky positions
and polarization angles for injections covering the entire S2 data set.
The $x$ axis reflects the $\hrss$ as defined in equation~\ref{eq:hrss}.
}
\label{fig:sg}
\end{figure}

\begin{figure}[!thb]
\includegraphics[width=0.98\linewidth]{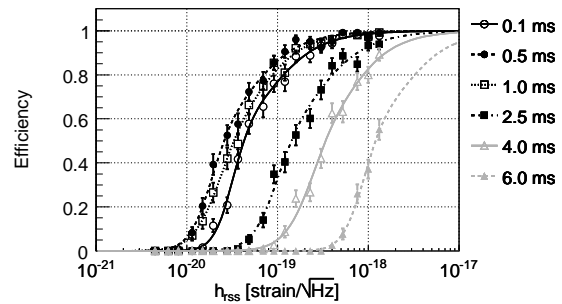}
\caption{Same plot as in Fig.~\ref{fig:sg} but
for Gaussian injections of $\tau=$0.1, 0.5, 1.0, 2.5, 4.0 and 6.0~ms.}
\label{fig:g}
\end{figure}

In Fig.~\ref{fig:sg} we show the efficiency curves, \ie, the efficiency
versus signal strength (at the Earth) 
of our end-to-end burst search pipeline for the
case of the six different sine-Gaussian waveforms we have introduced
earlier in this section.
As described in the previous subsection, these efficiency curves reflect
averaging over random sky positions and polarization angles.
As expected given the instruments' noise floor (see Fig.~\ref{fig:s2sens}),
the best
sensitivity is attained for sine-Gaussians with a central frequency
of 235~Hz; for this signal type, the required strength in order to reach 50\%
efficiency is $\hrss=$1.5$\times 10^{-20}$~\hrssu
, which is roughly a factor of 20 above the noise floor of
the least sensitive LIGO instrument at 235~Hz during S2.
In Fig.~\ref{fig:g} we show the same curves for the Gaussian family of
waveforms we considered.
The 6~ms Gaussian presents the worst sensitivity because most of
its signal power is below 100~Hz.
The maximum $\hrss$ used for the Gaussian injections was $1.32 \times
10^{-18}$~\hrssu; we cannot rely on the fitted curves to accurately
extrapolate the efficiencies much beyond that $\hrss$.
The sensitivity of this search to
$\hrss$ for these two families of waveforms
is summarized in Table~\ref{tbl:adhoc}.

\begin{table} [!thb]
\caption{Summary of the S2 pipeline $\hrss$
(from equation~\ref{eq:hrss})
sensitivity
to {\it{ad hoc}} waveforms
in units of $10^{-20}$~\hrssu.
These values are averages over random sky positions and signal
polarizations.  The injections did not span a wide enough range of
amplitudes to accurately determine the 95\% efficiency value for the
$\tau$=$4.0$~ms Gaussian, nor the 90\% and 95\% efficiency values for
the $\tau$=$6.0$~ms Gaussian.}
\label{tbl:adhoc}
\begin{tabular}{lccc}
\hline\hline
                          & 50\% & 90\% & 95\% \\
\hline
sine-Gaussian $f_0$=100~Hz &  8.2    & 33     &  53\\
sine-Gaussian $f_0$=153~Hz &  5.5    & 24     &  40\\
sine-Gaussian $f_0$=235~Hz &  1.5    & 7.6    &  13\\
sine-Gaussian $f_0$=361~Hz &  1.7    & 8.2    &  14\\
sine-Gaussian $f_0$=554~Hz &  2.3    & 10     &  17\\
sine-Gaussian $f_0$=849~Hz &  3.9    & 20     &  34\\
\hline\hline
Gaussian $\tau$=0.1~ms     &  4.3   & 21     &  37 \\
Gaussian $\tau$=0.5~ms     &  2.6   & 13     &  22 \\
Gaussian $\tau$=1.0~ms     &  3.3   & 16     &  26 \\
Gaussian $\tau$=2.5~ms     &  14    & 75     &  130\\
Gaussian $\tau$=4.0~ms     &  34    & 154    &  ---\\
Gaussian $\tau$=6.0~ms     &  121   & ---    &  ---\\
\hline\hline
\end{tabular}
\end{table}

\subsection{Signal parameter estimation}
The software signal injections we just described provide a good
way of not only measuring the efficiency of the search
but also benchmarking WaveBurst's ability to
extract the signal parameters.
An accurate estimation of the signal parameters by a detection
algorithm is essential for the successful use of
time and frequency coincidence among candidate triggers
coming from the three LIGO detectors.

\begin{figure}[!thb]
\includegraphics[width=0.98\linewidth]{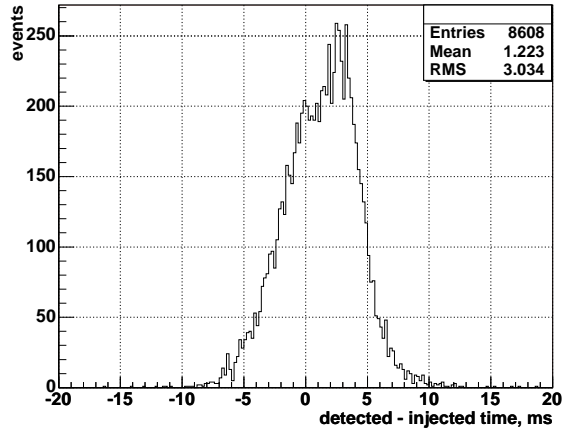}
\caption{Timing error of the WaveBurst algorithm in the three
LIGO instruments during S2 when sine-Gaussian injections of
varying frequency and strength were injected.
For comparison, the time separation between the two LIGO sites
is 10~ms and the coincidence time window used in this analysis
is 20~ms.}
\label{fig:sg_time}
\end{figure}

We compare the central time of a WaveBurst event
(section~\ref{sec:etg_methods}) with the known central time
of the signal injection.
For each of the two {\it{ad hoc}} waveform families considered so far,
as well
as for each of the astrophysical waveforms
we will discuss in section~\ref{sec:results},
WaveBurst is able to resolve
the time of the event on the average with a systematic shift of less
than 3~ms and with a standard deviation of the same value.
In Fig.~\ref{fig:sg_time} we show a typical plot of the timing error for
the case of all the sine-Gaussian injections we injected in
the software simulations and for the three LIGO instruments together.
The apparent deviation from zero has a contribution
coming from the calibration phase error.
Another contribution comes from the fact that the detected central
time is based on a finite time-frequency volume of the signal's
decomposition which is obtained after thresholding.
It remains however well within our needs for a tight time
coincidence between interferometers.
For the same type of signals, we list in Table~\ref{tbl:sg_freq}
the reconstructed versus injected central frequency.
The measurements are consistent within the signal bandwidth.

\begin{table} [!thb]
\caption{Central frequency reconstruction for Q=8.9 sine-Gaussian
injections.}
\label{tbl:sg_freq}
\begin{tabular}{c|c|c}
\hline\hline
Injected & Mean of detected & Standard deviation of  \\
frequency (Hz) & frequency (Hz) & detected frequency (Hz) \\
\hline
100 & 98.4 & 3.9 \\
153 & 159.5 & 4.4 \\
235 & 242.7 & 14.2 \\
361 & 363.7 & 14.0 \\
554 & 544.3 & 17.0 \\
849 & 844.9 & 21.4 \\
\hline\hline
\end{tabular}
\end{table}

\begin{figure}[!thb]
\includegraphics[width=0.98\linewidth]{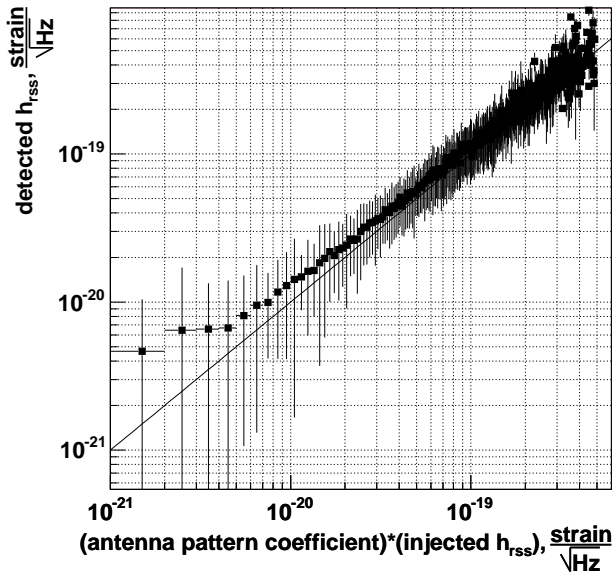}
\caption{Detected versus injected {\it root-sum-square} (rss)
signal strength
for Q=8.9 sine-Gaussian injections. The vertical bars
indicate one-sigma spread of the reconstructed values.
\label{fig:sg_hrss}}
\end{figure}

The WaveBurst algorithm estimates the signal strength from
the measured excess power in the cluster pixels, expressed as $\hrss$
as in equation~\ref{eq:hrss} but with the integrand being the
antenna-pattern-corrected $h_{det}(t)$ given by equation~\ref{eq:hdet}
rather than the intrinsic $h(t)$ of the gravitational wave.
Figure~\ref{fig:sg_hrss} shows that this quantity is slightly
overestimated on average, particularly for weak signals.
Several factors contribute to mis-estimation of the signal strength.
WaveBurst limits the signal $\hrss$
integration to within the detected time-frequency
volume of an event and not over the entire theoretical support
of a signal.
Errors in the determination of the signal's time-frequency
volume due to thresholding may lead to systematic uncertainties
in the determination of its strength.
The $\hrss$ shown in Fig.~\ref{fig:sg_hrss} also reflects
the folding of the measurements from all three
LIGO instruments and thus it is affected by
calibration errors and noise fluctuations in any instrument.
Our simulation analysis has shown that
the detected signal's $\hrss$ is the quantity most sensitive
to detector noise and its variability;
for this reason, it is not used in any step of the
analysis either as part of the coincidence analysis or for the
final event selection.

\subsection{Hardware injection results}
During the S2 data taking, as well as shortly before and after it, several
run intervals were designated for hardware signal injections.
These injections
were intended to address any instrumental issues, including calibrations,
and provide a robust end-to-end test of LIGO's data analysis pipelines.
They also provided an important tool for
establishing the ``safety'' of the veto analysis, \ie, 
the absence of any couplings between a real gravitational
wave burst and the auxiliary channels we considered as potential
vetoes (section~\ref{sec:vetoes}).
An arbitrary function generator connected to the 
mirror position actuators provided
the capability of exciting the mirrors according to a simulated
gravitational wave pattern.
The waveforms injected through this hardware calibration included several
of the ones described in the target waveform section above.
The signals were injected into all three LIGO instruments at identical
times, without attempting to mock up the relative time delays and
amplitudes that would be produced by a source at a particular position
in the sky.
Thus the coincidence analysis, using the end-to-end pipeline
invoked in the analysis of the real data alone as well as with
software injections, was not fully appropriate for the hardware injections.
We have restricted ourselves to examining the performance of the
LIGO instruments and of the WaveBurst ETG in detecting these events
and reconstructing their signal parameters using each individual
detector.
Both the time and frequency reconstruction by the WaveBurst algorithm
on these hardware-injected signals is consistent with our software
injections and within our expectations.

\subsection{Error analysis}

The largest source of systematic error in the 
efficiency of this search is uncertainty in the absolute
calibration of the detectors.  Several contributions to this
uncertainty have been considered~\cite{ref:S2calibration}.
Systematic uncertainties are less
than 12\% for
L1, 5\% for H1, and 6\% for H2 over the frequency band used in this
analysis.  The calibration at any given point in time is subject to an
additional uncertainty from detector noise affecting the measurement
of the amplitude of the calibration lines.  These random errors
were especially large near the beginning of the run, when the H1 and L1
calibration lines were rather weak.  However, the efficiency of the
search, averaged over the run, is insensitive to these random errors.
The overall systematic uncertainty on the triple-coincidence
efficiency is a combination
of the individual systematic uncertainties which depends on the
relative sensitivities of the detectors, with the least sensitive
detector having the greatest influence.  As shown in
Fig.~\ref{fig:s2sens}, H2 was the least sensitive detector at low
frequencies while H1 was the least sensitive at high frequencies.  The
net uncertainty in the efficiency is estimated to be less than 8\%
at all frequencies.

No significant systematic error is attributed to the procedure
we followed in order to perform the efficiency measurement.
The various signal morphologies were superimposed over 
the entire S2 data sample and its full range of detector behavior.
The statistical error attributed to the finite number of simulations
used for the efficiency measurement is reflected in
the goodness of the sigmoid-like fits and is estimated to be less
than 5\%.
The efficiency measurement was
performed in multiple slightly-varying ways all of which
yielded results within one standard deviation.
These variations included different sampling of the S2 data set,
different versions of the calibration constants, and
different number and placement of the signal injections.

Combining all uncertainties, we estimate our efficiency to
any given signal morphology to be accurate at the 10\% level
or better.

\section{Search Results}\label{sec:results}
\subsection{Rate versus strength upper limit}
As we have seen in section~\ref{sec:evana}, using the zero-lag and
background rate measurements we set an upper bound
on the rate of gravitational wave bursts at the instruments at the
level of $0.26$ events per day at the 90\% confidence level.
We will now use the measurement of the efficiency of the
search as described in the previous section
in order to associate the above rate bound with the strength of the
gravitational wave burst events.
This is the rate versus strength interpretation
that we introduced in our
previous search for bursts in LIGO using the S1 data~\cite{ref:BUL}.

\begin{figure}[!thb]
\includegraphics[width=0.98\linewidth]{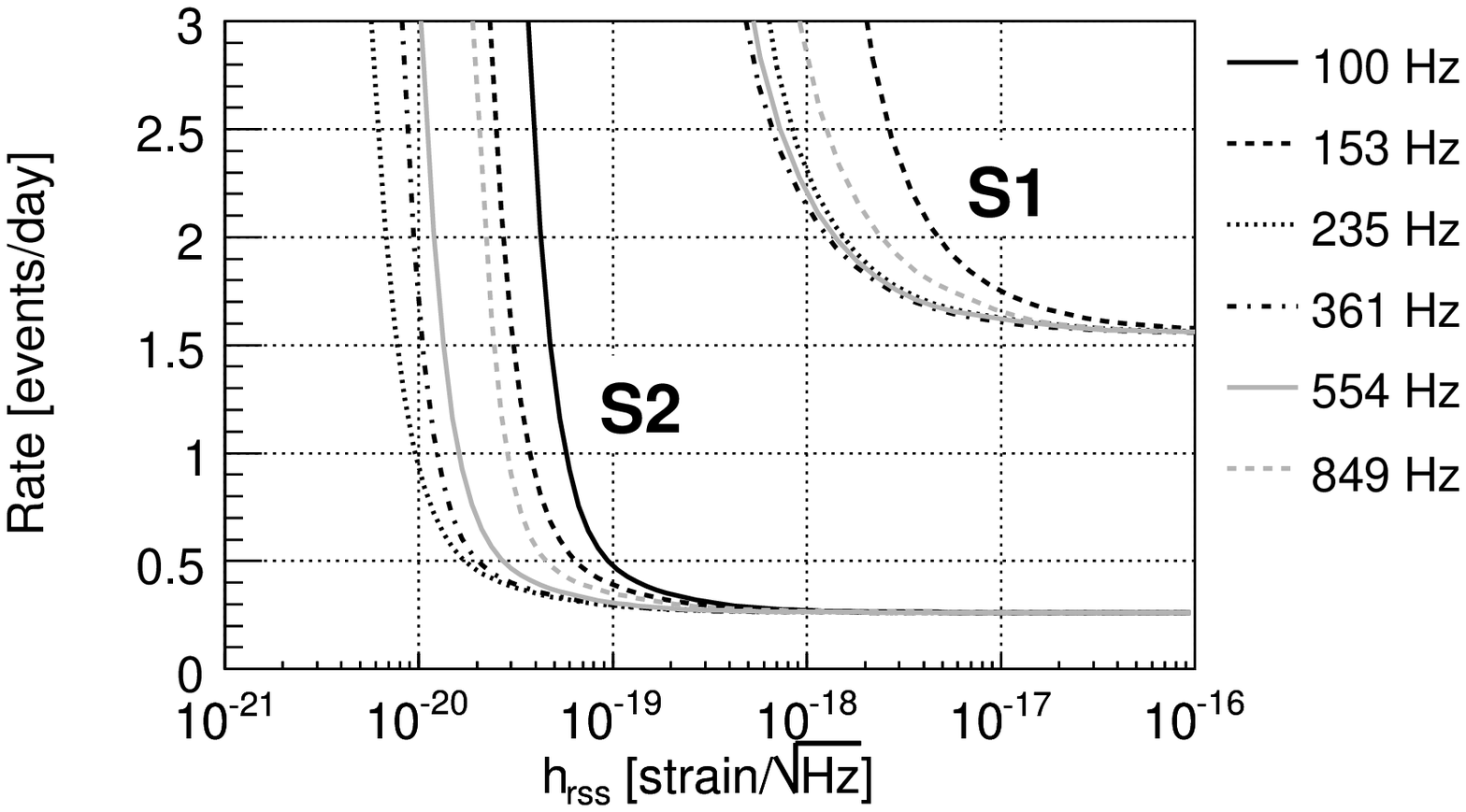}
\includegraphics[width=0.98\linewidth]{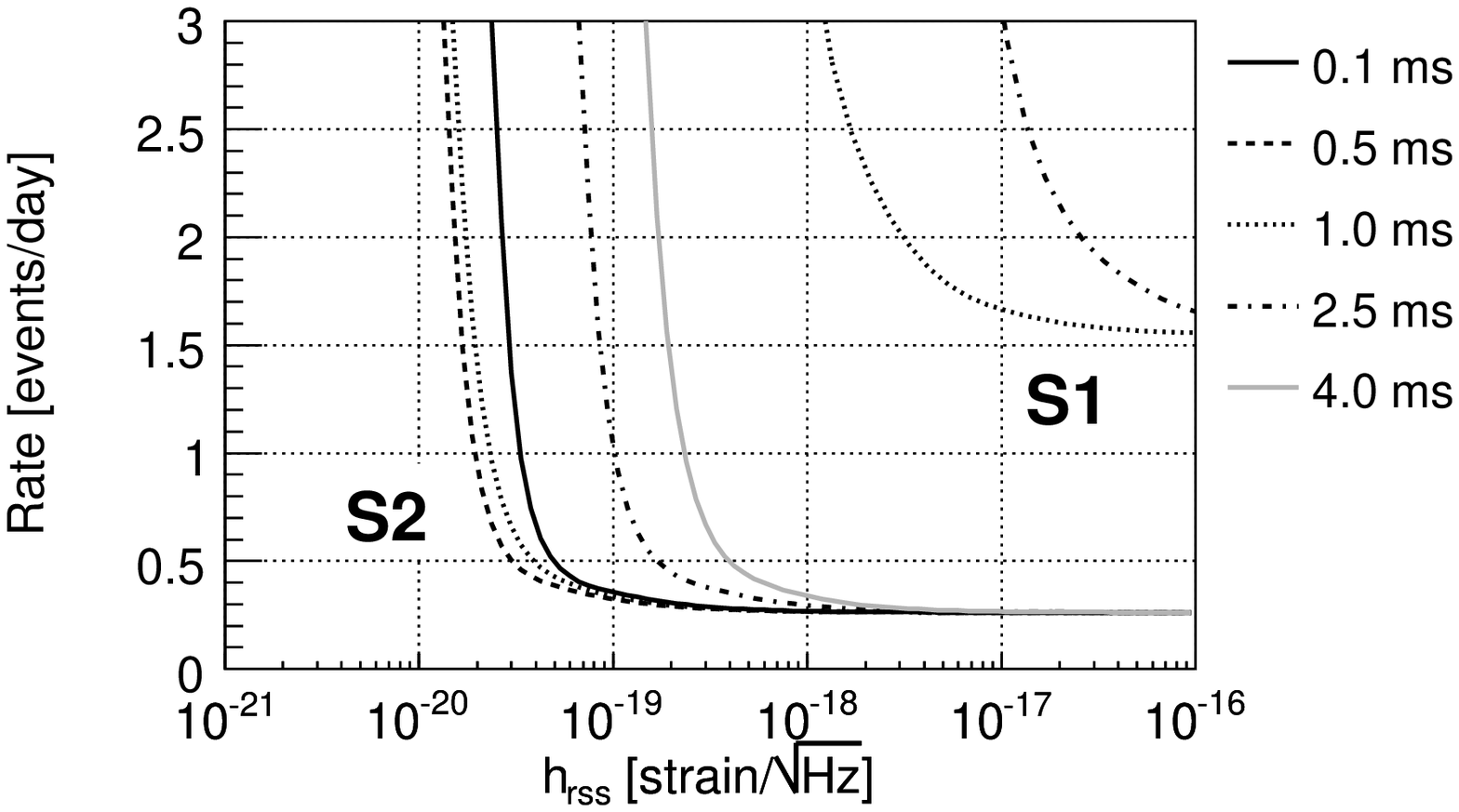}
\caption{Rate versus $\hrss$ exclusion plots at the 90\%
confidence level derived from
the LIGO burst search using the S2 data.
The top plot corresponds to 
burst events modeled
by sine-Gaussians of Q=8.9 and frequencies ranging from 100~Hz to
849~Hz, 
while 
the bottom plot corresponds to 
ones events modeled by Gaussians of the
$\tau$'s shown.
For comparison, the corresponding curves resulting from the
S1 analysis are superimposed.
}
\label{fig:gsg_exclude}
\end{figure}

The rate bound of our search as a function
of signal strength $\hrss$
is given by
\begin{equation}
R(\hrss)=\frac{\eta}{\epsilon(\hrss)}
\end{equation}
where the numerator $\eta$ is the
upper bound on the rate of detectable signal events at
a given confidence level (Sec.~\ref{ss:ratelimit})
and the denominator is the fractional
efficiency for signals of strength $\hrss$ (at the Earth).
This rate versus strength interpretation makes the same assumptions on the
signal morphology and origin as the ones that enter in the determination
of the efficiency.
In Fig.~\ref{fig:gsg_exclude} we show the rate versus strength upper
limit for the sine-Gaussian and Gaussian waveform families.
For a given signal strength $\hrss$ these plots give the upper limit
at the 90\% confidence level on the rate of burst events at the instruments
with strength equal to or greater than $\hrss$.
In that sense, the part of the plot above 
and to the right of these curves defines the region
of signal strength-rate
excluded by this search at 90\% confidence level.
As one would expect, for strong enough signals the efficiency of the search
is 1 for all the signal morphologies: this part of the plot remains
flat at a level that is set primarily by the observation time of this search.
For weaker signals the efficiency decreases and the strength-rate plot
curves up.
Eventually, as the efficiency vanishes the rate limit reaches
infinity asymptotically.
These curves for the various waveforms are not identical, as the detailed
trailing off of the efficiency is dependent on the waveform.
The exclusion rate-strength plots obtained from the S2 analysis represent
a significant improvement with respect to the S1 result~\cite{ref:BUL}.
As already noted in section~\ref{sec:evana}, the horizontal part of the
plot determined by the observation time is improved by a factor of 6
while the sensitivity-limited curved part of it reflects an improvement
in the efficiency of a factor of 17 or better, depending
on the waveform morphology.

\subsection{Astrophysical waveforms}

As mentioned in the introduction,
potential sources of gravitational wave bursts 
targeted in this search include core collapse supernovae,
merging compact binaries (neutron stars
and/or black holes) and gamma ray bursts.
In recent years there has been much effort
devoted to predicting gravitational wave burst waveforms from 
astrophysical sources,
generally relying on detailed numerical and approximation methods.
Our search is designed to be sensitive to a broad range of
short-duration bursts, so we wish to evaluate how it performs for
plausible astrophysical signals suggested by certain models.
As part of our signal simulation analysis for this search,
we focus in particular on the case of the core collapse
of rapidly spinning massive stars
~\cite{ref:ZM,ref:DFM,ref:OBLW}, 
and of binary black hole mergers~\cite{lazarus,ref:Laz,astrogravs}.

The core collapse simulations employ detailed hydrodynamical models
in two dimensions, enforcing axisymmetry of the rotating star
throughout its evolution. 
The core collapse is initiated artificially
(\eg, through a change in the adiabatic index of the
core material~\cite{ref:ZM}). 
An accelerating quadrupole moment 
is calculated in 2D from the distribution and flow of matter
during the collapse, 
from which the gravitational wave signal is derived.
The rapid spinning of the progenitor star may produce
multiple bounces of the dense core, which is reflected
in the waveform of the emitted waves.
Simple models of the differential rotation of material
in the star also lead to significant differences
in the resulting waveforms.
Relativistic effects~\cite{ref:DFM}, if included,
serve to effectively ``stiffen'' the core,
shifting the waves to higher frequencies and shorter durations.
The simulation is followed through the 
core collapse phase when most of the gravitational wave signal
is produced; it need not be continued through to the explosion
of the outer layers (and indeed, these simulations may not
produce such explosions).
The simulations attempt to sample the space
of important parameters (progenitor star angular momentum,
differential angular momentum versus radius,
density versus radius, adiabatic index of the core, \etc),
resulting in collections of waveforms with widely varying
morphologies; 
but of course the actual distributions of such parameters
are poorly known. 
In ref.~\cite{ref:OBLW} the authors employ updated
progenitor models and nuclear equation of state.
For the studies described here, we make use of 
78 waveforms supplied in ref.~\cite{ref:ZM},
26 from ref.~\cite{ref:DFM},
and 72 from ref.~\cite{ref:OBLW}.
We emphasize that we are studying these waveforms 
only as a guide for evaluating our search algorithm; 
we do not rely on accurately modelling a realistic population of
progenitor stars.

The binary black hole merger waveforms are taken from 
the Lazarus project~\cite{lazarus,ref:Laz,astrogravs}, which combines numerical
simulation 
of the vacuum Einstein equations for the most significantly 
nonlinear part of the interaction with close-limit perturbation 
theory for the late-time dynamics.
The authors in~\cite{lazarus,ref:Laz,astrogravs} generate 
waveforms from simulations of equal mass binary
black holes with no intrinsic spin starting from near
the innermost stable circular orbit following
a binary black hole inspiral.
It should be kept in mind that these waveforms 
include the ringdown phase of the binary
system and would naturally
occur after an inspiral waveform, which is searched for using
matched filtering techniques~\cite{ref:IUL,ref:IUL2}.

In all of these models, the simulations and calculations predict 
gravitational wave bursts with time durations 
ranging from a fraction of a millisecond 
to tens or hundreds of milliseconds and with a significant fraction of their
power in LIGO's most sensitive frequency band (100--1100~Hz).
This observation motivates the choice of parameters for the sine-Gaussian and
Gaussian waveforms used to optimize and evaluate
the efficiency for our search pipeline, as discussed in
section~\ref{sec:efficiency}.
After tuning our pipeline algorithms using these {\it{ad hoc}}
waveforms, we evaluate the efficiency for our search 
to detect the waveforms predicted from the astrophysical simulations.
The amplitudes of these waveforms are predicted by the simulations,
so that in addition to evaluating the efficiency as a function of
$\hrss$ (at the Earth), we can also present them as a function
of distance to the source (for a particular source inclination).
We have evaluated the efficiency versus actual distance,
averaged over source directions and polarizations,
assuming an isotropic distribution in source direction.
This assumption becomes invalid for supernova progenitors in
the Galactic disk, when the LIGO detectors become
sensitive to supernovae at distances greater than the 
disk thickness (on the order of 150~pc).
It is also invalid for extra-galactic binary black hole mergers,
since the distribution of nearby galaxies is far from
isotropic at the 1~Mpc scale.
This evaluation of the efficiency as a function
of distance is only to ``set the scale'' for 
the current and future astrophysical searches.

\begin{figure}[!thb]
\includegraphics[width=0.98\linewidth]{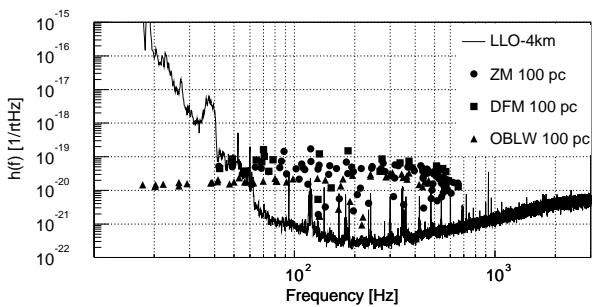}
\caption{Signal strength $\hrss$ at the detectors versus
central frequency for the 176 supernovae waveforms from
the three models described in references~\cite{ref:ZM,ref:DFM,ref:OBLW}:
the hydrodynamical model of ref.~\cite{ref:ZM}, labeled ``ZM'',
the relativistic effects considered in ref.~\cite{ref:DFM} and labeled ``DFM'',
and finally the hydrodynamical model employing realistic
nuclear equation of state of ref.~\cite{ref:OBLW}, labeled ``OBLW''.
In all cases,
the supernova events are positioned in optimal orientation and
polarization at 100~pc from the detectors.
The strain sensitivity of the L1 detector during the S2 run is shown
for comparison.
\label{fig:SN2}}
\end{figure}

For the case of core collapse supernovae we considered
the collections of waveforms from the three studies
discussed above~\cite{ref:ZM,ref:DFM,ref:OBLW}.
There are 176 such supernovae waveforms.
They are generally broad-band in frequency;
for 115 of them their central frequency is within
the sensitive band of this search (100--1100~Hz)
and for them we 
established strength and distance sensitivities.
Sources were uniformly
distributed over the whole sky with random polarization 
and fixed, optimal inclination.
For detecting these waveforms,  $\hrss$
amplitudes of a few times $10^{-20}$~\hrssu\
corresponding to source distances of the order of 100 pc were
required.
Such close-by supernovae are, of course, quite rare.
In Fig.~\ref{fig:SN2} we show the expected $\hrss$ (at the detectors)
and the central frequency 
for each of the 176 supernovae waveforms
assuming they originate from a core collapse supernovae that is
optimally oriented and polarized and is located at 100~pc from the detectors.
Superimposed, we show the sensitivity of the LIGO instruments during S2.

For the case of binary black hole mergers, we have
considered systems of black hole mass in the range of
10-90~$M_\odot$.
The characteristic frequency of the resulting waveform is
inversely proportional
to the mass of the system and thus five different masses of 10, 30, 50,
70 and 90~$M_\odot$  were chosen in order to span the nominal
frequency band of this search, \ie, the 100--1100~Hz band.
Moreover, the waveforms of these systems~\cite{lazarus,ref:Laz,astrogravs}
come with two polarizations
and they thus offered a check of the robustness of the
waveform consistency test, the $r$-statistic, against complex
morphologies.
The efficiency is calculated over the whole sky considering
the two polarization waveforms and a fixed inclination angle.
The best performing mass system corresponds to 50~$M_\odot$:
the characteristic frequency of this system corresponds to the best operating
point of the LIGO instruments, \ie, close to 250~Hz.
On the contrary, the two worst performing mass systems reflect 
frequencies at the two ends of the LIGO instrument's sensitivities
relevant to this search, \ie, 100~Hz and 1100~Hz.
As with the supernova waveforms, the black-hole simulations provide us with 
order-of-magnitude estimates of the distance to which our detectors 
were sensitive to such astrophysical systems during the S2 run.  
For the Lazarus black-hole mergers our $\hrss$
amplitudes of a few times $10^{-20}$~\hrssu\
correspond to 
distances of order 1~Mpc.

All four waveform families we have considered
for our simulations, either {\it{ad hoc}} or astrophysically motivated,
have frequency content that ranges over the entire band of our search.
Within each of these families, signal strengths in order to reach fixed
efficiency (\eg, 50\%) range over approximately an order of magnitude;
this is primarily a manifestation of the different frequency content of
each waveform and the fact that the LIGO detectors' frequency response
is not flat (and varies by an order of magnitude or more
over the 100--1100~Hz band, as shown in Fig.~\ref{fig:s2sens}).
We are in the process of augmenting the waveform library to be considered
in future LIGO burst searches.
This will give us further
opportunities to test the robustness of our methods and the use of
$\hrss$ as a measure
of the signal strength relevant to burst detection in LIGO.

\section{Summary and Discussion}\label{sec:summaries}

We have presented a search for gravitational wave bursts using the
data the three LIGO detectors collected during their second science run.
Transients with sufficient energy in LIGO's most sensitive
band during the S2 run, 100--1100~Hz, were searched for.
A search for gravitational wave bursts with frequency content above
the 1100~Hz range is being pursued in coincidence with the
TAMA~\cite{ref:TAMAref}
detector and will be described in a separate publication~\cite{ref:tamaligos2}.
Our analysis yielded a single candidate event which was subsequently
determined to be terrestrial in origin and was vetoed retroactively.
Incorporating this into
a frequentist statistical approach, 
we set an upper limit on the rate of strong gravitational wave bursts
of $0.26$ events per day at the 90\% confidence level.
This rate limit is a factor of 6 below our previously published
value~\cite{ref:BUL}, due primarily to the longer duration of the S2 run
and to better data quality.

The efficiency of this search was measured using various waveform
morphologies.
Besides the families of {\it{ad hoc}} waveforms we introduced in
our previous search, we also measured the efficiency of our search
to astrophysically motivated waveforms resulting from numerical
simulations of the core collapse supernovae and binary black hole
mergers.
For most of the waveforms considered, the values of $\hrss$ at 50\%
efficiency lie in the
$10^{-20}-10^{-19}$~\hrssu\ range.
The sensitivity attained by this search represents an improvement
with respect to S1 by a factor of 17 or more for waveforms studied
in both searches.
This difference is frequency dependent and mainly reflects
the instruments' noise floor improvement by a factor of $\sim$10.
The rest is attributed to improvements of the search algorithm
and the use of the waveform consistency test ($r$-statistic),
allowing a lower effective threshold on signal amplitude.
The interpreted results of this search include the rate versus
strength exclusion curves on a per waveform morphology basis.
The improvements on the rate and signal strength sensitivity are
both reflected in significantly more stringent regions now allowed
in these rate versus strength plots.

\subsection{Comparison with previous searches}

In our S1 paper, we made a comparison with results from
searches with broad-band interferometric detectors described
in~\cite{ref:UGMPQ,ref:Forward}.
The upper limit set by these detectors is at the 
level of 0.94 events per day and with a signal strength
sensitivity of $\hrss = 5.9\times 10^{-18}$~\hrssu,
both of which are now surpassed by our S2 search.
In our S1 paper we also compared with the results of the
IGEC search for gravitational wave bursts~\cite{ref:IGEC2003}.
LIGO's broad band response allowed us to set better limits on bursts whose
power was mainly at frequencies away from the bars' resonances. At or near
the bars' resonant frequencies, however, the IGEC search benefited from a
much longer observing time and somewhat better sensitivity at those
frequencies, and thus was able to set rate limits far below what we were
able to do in LIGO. With improved sensitivity in S2, it is interesting to
again compare LIGO's performance at a frequency near the bars' resonance.
In order to perform this comparison in our published S1 work~\cite{ref:BUL}
we chose
the sine-Gaussian simulations at 849~Hz frequency and with Q=8.9.
Although this waveform morphology has significant signal power in the narrow
frequency band (895--930~Hz) of sensitivity for most of the IGEC detectors,
it actually fails to maintain an approximately flat Fourier spectrum
over the broader range (694--930~Hz) needed in order to encompass all of them.
For this reason,
in order to perform the same comparison in S2 we have used the
Gaussian of $\tau$=0.1~ms signal morphology which was the
closest waveform with a flat spectrum in the 694--930~Hz range
that we included in our S2 simulations.

\begin{figure}[!thb]
\includegraphics[width=1.05\linewidth]{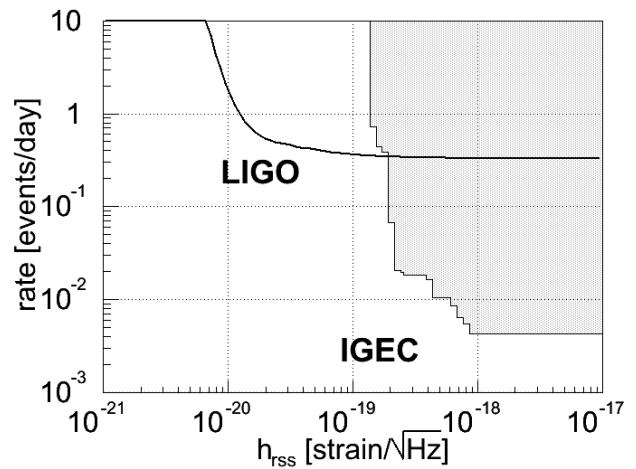}
\caption{Rate versus $\hrss$ exclusion curves at the 95\%
confidence level for optimally oriented 
Gaussians of $\tau$=0.1~ms.
The solid curve displays the 95\% confidence level 
measurement obtained by LIGO with this search.
The IGEC exclusion region is shown shaded and it is adapted
from Fig.~13 of~\cite{ref:IGEC2003}.
If the comparison were performed using
Q=8.9, 849~Hz sine-Gaussians, the LIGO and IGEC curves
would move to smaller amplitudes by factors of 
1.1 and $\sim$3, respectively.
}
\label{fig:LIGOvsIGEC}
\end{figure}

The IGEC analysis~\cite{ref:IGEC2003}
set an upper limit of $\sim 4\times 10^{-3}$ events/day
at the 95\% confidence level on the rate of gravitational wave bursts.
The limit was derived assuming optimal source direction and
polarization
and was also given as a function of the burst Fourier amplitude
in a rate versus strength exclusion curve similar to LIGO's.
Using LIGO's S2 Gaussian of $\tau$=0.1~ms simulations, the
WaveBurst ETG efficiency for sources with random linear polarizations is
50\% at a strength of $4.3 \times 10^{-20}$~\hrssu\
(see Table~\ref{tbl:adhoc}). 
For the same sources all with {\it optimal}
polarizations, the 50\% efficiency point improves by roughly a
factor of 3, to $1.6 \times 10^{-20}$~\hrssu. Thus the
optimally oriented rate versus strength curve looks similar to
Fig.~\ref{fig:gsg_exclude}, but shifted to the left. Substituting
the 95\% confidence level (CL) event limit of $0.33$ for the 90\% CL event
limit of $0.26$ shifts
the curve up. Lastly, the IGEC excluded region from Fig.~13
of~\cite{ref:IGEC2003} can be translated from bars' natural units
(Hz$^{-1}$) to units of $\hrss$ (\hrssu).
Given the Fourier transform h(f) for a Gaussian waveform h(t),
\begin{equation}\label{eq:gft1}
  h(t)=\hrss\left(\frac{2}{\pi\tau^2}\right)^{1/4}\exp{(-t^2/\tau^2)}
\end{equation}
\begin{equation}\label{eq:gft2}
  h(f)=\hrss\left(2\pi\tau^2\right)^{1/4}\exp{(-\pi^2\tau^2f^2)},
\end{equation}
we convert the IGEC values of spectral amplitude h(f) into $\hrss$
for a Gaussian of $\tau$=0.1~ms signal morphology (the conversion is
a function of the assumed frequency of the IGEC result and may vary
by a few percent over the 694--930~Hz range.)
The resulting comparison can be seen in Fig.~\ref{fig:LIGOvsIGEC}.

With LIGO and S2 we are able to stretch the excluded region substantially
to the left (\ie, to weaker signals) 
of the boundary of the excluded region in the rate versus strength
curve of Fig.~13 in~\cite{ref:IGEC2003}. However, although S2's increased
observing time allows a lower rate limit than could be set in S1, it is
still the case that the IGEC longer duration search allows substantially better rate
limits to be set, for signals strong enough for its detectors to have
seen them.

Furthermore, we are interested in a comparison with the results reported
from the analysis of the EXPLORER and NAUTILUS 2001 data.
In their 2002 paper~\cite{rome2001} the Rome group that analyzes
the data of and operates these
two resonant mass detectors reported a
slight excess of events seen in sidereal hours between 3 and 5.
The events seen in concidence by the two detectors are of
an average temperature of approximately 120mK which
according to the authors corresponds to an optimally oriented
gravitational wave burst  Fourier amplitude of
$2.7 \times 10^{-21}$Hz$^{-1}$ (equation 4 from ref.~\cite{rome2001}).
The rate of such events is of order 
200 events/year (or 0.55 events/day)~\cite{rome2001,coccia2004}.
Given the amplitude of the observed events by the resonant mass
detectors, the corresponding $\hrss$ of the hypothetical
events in our LIGO instruments will generally depend
on the signal morphology.
As with our aforementioned IGEC analysis,
we considered the case of a Gaussian of $\tau$=0.1~ms,
for which the Fourier amplitude of the observed events
at the detectors' average resonance frequency implies
an $\hrss$ of $1.9\times 10^{-19}$~\hrssu.
Keeping
in mind that the suggested $\hrss$ values refer to optimal
orientation, we can see from Fig.~\ref{fig:LIGOvsIGEC} that
for this event strength the LIGO S2 search set an upper bound to
their flux at roughly 
$0.4$ events per day at the 95\% CL.
It should be noted, though, that depending on the assumptions of
signal waveform (for example a single cycle of a 914~Hz sinusoid
or a narrow-band sine-Gaussian signal centered on the same frequency)
or considering the range of event strengths recorded by
EXPLORER and NAUTILUS (rather than
their average value only) the corresponding $\hrss$ at the
LIGO detectors may come nearer to the threshold of our sensitivity
and thus make our rate limits poorer.
The signal strength and rate of the 2001 Rome results come with
enough uncertainties that given the LIGO S2 sensitivity and exposure
we cannot make a definitive comparison.
The significant 
improvements in sensitivity and longer observation times
that we expect in new LIGO searches in the near future will 
enable us to move in this direction.

\subsection{Discussion and future directions}

The search for gravitational wave bursts in LIGO's S2 run has
seen significant improvements introduced in the search methodology
and interpretation with respect to S1.
This included the introduction of the waveform consistency cut
and the use of astrophysically motivated waveforms as part of the search
interpretation.
Additional improvements are currently under way. We expect them
to bring stronger suppression of the background via the use of
a burst amplitude consistency test between the LIGO detectors as well
as new ways of performing our event analysis within the context
of a distributional analysis of their strength.
Moreover, we plan to make use of data taking periods corresponding
to the double coincidence of
the instruments that are not part of the triple coincidence dataset.
We will continue investigating the optimization of search algorithms
for specific types of waveforms and adding stronger astrophysical
context in our search by invoking source population models or targeting
plausible point sources.
Among the lessons learned in the S2 search has been the importance of
the follow-up investigations dictated by coincident triggers
revealed by the pipeline.
As expected, with the detector performance nearing design sensitivity,
potential couplings from the environment and the instrument
itself will become apparent and will need to be identified
and vetoed out.
Our on-going work for 
vetoes will become
more prominent together with the need to define rigorous criteria
and procedures for following up on such events.
LIGO's subsequent runs have already collected data of comparable duration
and improved sensitivity with respect to S2 and they will present
the next milestone of the search for bursts where a good number of these
improvements will be exercised.

The software used in this analysis is available in the
LIGO Scientific Collaboration's CVS archives at
{\tt{
http://www.lsc-group.phys.uwm.edu/cgi-bin/}} 
{\tt{cvs/viewcvs.cgi/?cvsroot=lscsoft}}
under the {\tt{S2\_072704}} tag for WaveBurst in {\tt{LAL}} and {\tt{LALWRAPPER}}
and {\tt{rStat-1-2}} tag for $r$-statistic in {\tt{MATAPPS}}.
\begin{acknowledgments}\label{sec:acknow}

The authors gratefully acknowledge the support of the United States National 
Science Foundation for the construction and operation of the LIGO Laboratory 
and the Particle Physics and Astronomy Research Council of the United Kingdom, 
the Max-Planck-Society and the State of Niedersachsen/Germany for support of 
the construction and operation of the GEO600 detector. The authors also 
gratefully acknowledge the support of the research by these agencies and by the
Australian Research Council, the Natural Sciences and Engineering Research 
Council of Canada, the Council of Scientific and Industrial Research of India, 
the Department of Science and Technology of India, the Spanish Ministerio de 
Educacion y Ciencia, the John Simon Guggenheim Foundation, the Leverhulme
Trust, the David and 
Lucile Packard Foundation, the Research Corporation, and the Alfred P. Sloan 
Foundation.

This document has been assigned LIGO Laboratory document number 
LIGO-\ligodoc.
\end{acknowledgments}
\bibliographystyle{apsrev}

\end{document}